\documentclass[12pt,preprint]{aastex}
\usepackage{rotating}

\newcommand{\myemail}{virgilif@physics.unlv.edu \\ zhang@physics.unlv.edu}

\shorttitle{Are all short-hard bursts mergers?}
\shortauthors{Virgili et al.}

\begin{document}


\title{Are all short-hard gamma-ray bursts produced from mergers of 
compact stellar objects?}

\author{Francisco J. Virgili and Bing Zhang}
\affil{Department of Physics and Astronomy, University of Nevada Las Vegas,
    Las Vegas, NV 89154, USA}

\email{\myemail}

\author{Paul O'Brien}
\affil{Department of Physics and Astronomy, University of Leicester, Leicester, LE1 7RH, UK}

\and

\author{Eleonora Troja}
\affil{NASA Postdoctoral Program Fellow, Goddard Space Flight Center, Greenbelt, MD 20771, USA}

\begin{abstract}

The origin and progenitors of short-hard gamma-ray bursts remain 
a puzzle and a highly debated topic. Recent Swift observations
suggest that these GRBs may be related to catastrophic explosions 
in degenerate compact stars, denoted as ``Type I'' GRBs. The most
popular models include the merger of two compact stellar objects 
(NS-NS or NS-BH).
We utilize a Monte Carlo approach to determine whether a
merger progenitor model can self-consistently account for all the 
observations of short-hard GRBs, including a sample with redshift
measurements in the Swift era ($z$-known sample) and the CGRO/BATSE 
sample. We apply various merger time delay distributions invoked
in compact star merger models to derive the redshift distributions 
of these Type I GRBs, and then constrain the unknown luminosity 
function of Type I GRBs using the observed luminosity-redshift ($L-z$)
distributions of the $z$-known sample. The best luminosity function
model, together with the adopted merger delay model, are then applied
to confront the peak flux distribution ($\log N - \log P$ 
distribution) of the BATSE and Swift samples. We find that 
for all the merger models invoking a range of merger delay time
scales (including those invoking a large fraction of ``prompt mergers''), 
it is difficult to reconcile the models with all the data. The data
are instead statistically consistent with the following 
two possible scenarios.  First, that short/hard GRBs are a superposition 
of compact-star-merger-origin (Type I) GRBs and a population of 
GRBs that track the star formation history,
which are probably related to the deaths of massive stars (Type II GRBs).
Second, the entire short/hard GRB population is consistent with a 
typical delay of 2 Gyr with respect to the star formation history with 
modest scatter. This may point towards a different Type I progenitor
than the traditional compact star merger models.
\end{abstract}

\keywords{gamma rays: bursts --- gamma rays: observations --- methods: statistical}

\section{Introduction}

Short-hard gamma ray bursts (GRBs) have been an enigma since the
identification of a bimodal distribution in the CRGO/BATSE data by
Kouveliotou et al. (1993).  This study showed that GRBs are
distributed into two populations with short (shorter than 2s) 
bursts having a harder spectrum and long (longer than 2s) 
bursts a softer spectrum, leading to the
short-hard/long-soft classification.  This purely observational
division has a fair amount of scatter and does not necessarily
indicate the nature of the intrinsic progenitor of a burst. 
Progress in understanding the progenitors of both long and short
GRBs was made following the discoveries of their respective afterglows
and host galaxies. While long bursts have been more securely shown to 
be associated with the
collapse of massive stars \cite{hjorth03,stanek03,campana06,pian06},
the identification of a progenitor type for short-hard bursts has not
been as successful.  The most popular model is a merger event between
two compact stellar objects, be it two neutron stars (NS-NS) or a NS and
a black hole (NS-BH)
\cite{lattimer76,paczynski86,eichler89,narayan92}.  
This is supported by
observational evidence of a \textit{lack} of a supernova component in
a handful of short-hard GRBs to deep limits
\cite{hjorth05,covino06,kann08} as well as the very important
discovery of a handful of short bursts identified in non-star
forming galaxies, such as GRBs 050509B and 050724, or at the
edge of star forming galaxies, such as GRB 050709
\cite{gehrels05,bloom06,fox05,villasenor05,hjorth05,barthelmy05,berger05}.
Such observational evidence, however, is not ubiquitous for all
short GRBs. In fact, most short-hard GRBs discovered later
are found in star forming galaxies or have missing hosts 
\cite{berger09}.

Prompted by the discovery of GRB 060614, a nearby long GRB without 
an associated supernova but with many properties consistent with
a merger-type progenitor
\cite{gehrels06,galyam06,fynbo06,dellavalle06},
Zhang et al. (2007) suggest that the long vs. short classification
of GRBs does not necessarily match the physical origins of 
massive star core collapse and mergers of compact stellar objects,
respectively, and one needs multiple observational criteria to
make correct identifications. 
The need of applying multiple criteria to determine
the physical origin of a GRB was also discussed by Donaghy et al. 
(2006). Zhang et al. (2007) suggest naming GRBs with massive star progenitor
and compact stellar object progenitor origins as ``Type II'' and ``Type I'',
respectively, so as to be differentiated from the traditional ``long''
and ``short'' terminology.  A more elaborate physical 
classification scheme was discussed by Bloom et al. (2008).
The multiple criteria to define Type II/I populations were elaborated 
in Zhang et al. (2009). They showed 
that not only could some long GRBs (such as GRB 060614) be of a
Type I origin, but also that a good fraction (not only the Gaussian 
tail of the long population) of short GRBs could be of a Type II 
origin. They argued that the two recently discovered high-$z$
GRBs with intrinsic short durations, GRBs 080913 and 090423, are
most likely Type II bursts, and further suggested that some
high-$z$, high-$L$ short-hard GRBs can be of Type II origin
as well. The goal of this work is to 
investigate through statistical methods whether the data are 
consistent with the hypothesis that ``all short/hard
GRBs are Type I'', in particular, whether they are related to 
compact star mergers.

Many of the specifics of Type I bursts are loosely constrained.  
Two important properties, among others, are the form of the 
luminosity function and the distribution of the merger delay time
scale $\tau$, which is defined as the time elapsed
between star formation and the GRB. For compact star merger models,
this is the delay between the formation of the two main sequence stars 
(i.e. the epoch of star formation) and the coalescence between the two
evolved compact stellar objects (NS-NS or NS-BH).  
Several studies have endeavored to
add constraints to these distributions
\cite{ando04,gandp05,galyam05,ngf06,gandp06}. 
For the merger delay time scale, usually a long delay
is invoked, in the form of either
a roughly constant delay
(anywhere from 1-6 Gyr) or a distribution that is proportional to 
a power $\gamma$ of the delay time scale $\tau$.  Nakar, Gal-Yam \& Fox
(2006b) constrained the delay distribution of merger events to $\tau > 4$
Gyr or a distribution $\propto \tau^{-0.5}$ or shallower, while 
Guetta and Piran (2006) concluded that this distribution can be 
modeled by a logarithmic delay or one with a constant delay, 
generally on the order of a few Gyr. Later, some short GRBs
with much higher redshifts were identified
\cite{berger07a,graham09}, which posed a challenge
to the models invoking a long merger delay time scale. 
A more physical approach is to model the delay time scale
through population synthesis 
\citep[e.g.][]{bandk01,bel02,ivanova03,dandp03,bel06}. 
These authors suggest that merger
timescales should not only be concentrated to long ``classical''
timescales \cite{bandv91} but also include a prompt merger channel.
These arguments stem from the details of the binary evolution process.
Belczynski and Kalogera (2001) as well as Ivanova et al. (2003) and
Dewi and Pols (2003) proposed a scenario where ultra-compact orbits
can be achieved by an extra mass transfer event in the evolution of
the binary, further reducing the orbital size of the final system that
produces the bursts.  This can lead to explosions 
on the order of 10s of Myr. This scenario is broadly consistent 
with the fact that short GRBs are seen 
in both early type galaxies and star forming galaxies
\cite{bel06,zheng07,zhang09}. However, it is unclear whether
the model can reproduce all the available data of short-hard GRBs.

The luminosity function of Type I GRBs is even more sparsely constrained.
Nakar et al. (2006b) assumed a simple powerlaw luminosity function
and found that an index -2 can fit the available data by the end 
of 2005. Guetta \& Piran (2006) introduced a broken powerlaw, with 
the indices in low luminosity $\sim -0.5$ and in high luminosity
ranging from -1 to -2.  Both
works also included the caveat that the observational sample is very
small, and that the data allows for some flexibility when combining
rates, luminosities, and delay distributions.  Modifications can also
be added by considerations of multiple populations of short-hard
bursts, such as a dual-peak luminosity function to account for local
SGR giant flare events \cite{tanvir05,chapman09}, or the contributions
of Type I GRBs from globular clusters \cite{grindlay06,salvaterra08},
but these contaminations are either not significant (Nakar et al. 
2006a) or without robust evidence for their existence.

Since the early attempts of constraining compact star merger progenitor
models \cite{ngf06,gandp06} shortly after the
discovery of the short GRB afterglows, the sample of 
short-hard GRBs with redshift measurements has significantly
expanded. Observations after 2005 suggest that nearby,
early-type galaxies are not common short GRB hosts, and that
a significant fraction of short GRBs are likely from the high
redshift universe \cite{berger07,berger09}. Zhang et al. 
(2009) argued that the
Type I Gold Sample, a small group of GRBs that carry direct
evidence in favor of a compact stellar object origin, are not
representative of the BATSE short-hard GRBs. In particular,
four out of the five GRBs in the sample have extended emission, and
all five have a moderate hardness ratio. Even without accounting
for the extended emission, the ``short spike'' of
GRB 050724 has a duration longer than 2 seconds. This GRB
is the strongest evidence for the compact star merger
origin of short GRBs to date, since its afterglow is within a nearby
early type galaxy \cite{barthelmy05,berger05}. After 
5 years of observations, however, this burst is still the only one with 
such a robust association\footnote{GRB 050509B \cite{gehrels05,bloom06}
is believed to be associated with a cD elliptical galaxy in a
nearby cluster. However, the argument was based on a chance
coincidence argument with the XRT error box, since no optical
afterglow was detected for this burst.}. Zhang et al. (2009)
suspect that some (maybe even most) short-hard GRBs are not
related to compact star mergers (Type I), but are related to 
deaths of massive stars (Type II). Recently, short GRB 090426 
was discovered at $z=2.6$. Several groups drew the conclusion that
this burst is likely of a Type II origin based on some
independent arguments (Levesque et al. 2009; Antonelli et al. 2009; 
Xin et al. 2010; L\"u et al. 2010).
Another related study is by Nysewander et al. 
(2009) (see also Kann et al. 2007, Kann et al. 2008), who 
analyzed the optical afterglow properties of both long and
short GRBs and found that the two populations have very similar
optical-to-X-ray flux ratios, suggesting that the average 
circumburst density of the two populations is similar. All these
facts call for a serious re-investigation of the hypothesis that
all short GRBs are related compact star mergers.

The current sample of short-hard GRBs with redshift measurements
(the $z$-known sample) is large enough to serve the purpose of
constraining redshift distribution and luminosity function
of the population. This sample, together with the BATSE and Swift
all short-hard GRB samples, can be used to perform a self-consistent
check of the hypothesis that ``all short-hard GRBs have compact
binary merger progenitors''. This is the goal of this 
paper. Assuming that all the $z$-known short/hard GRBs detected
in the Swift era are Type I GRBs, we perform a series of Monte 
Carlo simulations to constrain the luminosity function of these 
putative Type I GRBs by adopting
several time delay distribution models, and check whether
the same model can reproduce the peak flux distribution 
($\log N - \log P$) of the BATSE and Swift samples. 
In \S 2 we detail the model assumptions and some information on 
the simulations. In \S 3 we present the results of the constraints
on various models with different combinations of merger time 
delay distribution and luminosity function. Our results are
summarized in \S 4 with a short discussion.

\section{Models and Theoretical Framework}

An advantage of utilizing numerical methods to approach a problem is
that it can be easily broken down into its constituent parts for easy
processing.  The number of any type of GRBs that occur within a comoving
volume element, $dV/dz$, per unit observed time at redshift $z \sim z
+ dz$ and luminosity $L \sim L+dL$ is given by
\begin{equation}\label{dn}
\frac{dN}{dtdzdL}=\frac{R_{GRB}(z)}{1+z}\frac{dV(z)}{dz}\Phi(L),
\end{equation}
with the factor of $(1+z)$ accounting for the cosmological time
dilation, $R_{GRB}(z)$ being the GRB volume event rate (in unit of 
${\rm Gpc^{-3}~yr^{-1}}$) as a function of $z$, $\Phi(L)$
the luminosity function, and $dV(z)/dz$ the comoving volume element
given by
\begin{equation}\label{volume}
\frac{dV(z)}{dz}=\frac{c}{H_{\rm 0}}\frac{4\pi D_{L}^2}{(1+z)^2
[\Omega_M(1+z)^3+\Omega_\Lambda]^{1/2}},
\end{equation}
for a flat $\Lambda$ cold dark matter ($\Lambda$CDM) universe.
Throughout the work, $\Omega_m$ and $\Omega_\Lambda$ are set to 0.3
and 0.7, respectively. Both the expressions for the GRB volume
event rate ($R_{GRB}(z)$) and the luminosity
function ($\Phi(L)$) of Type I bursts are unknown, and are the major
focus of this study.  

First, there is no theoretical prediction on the form of the luminosity
function of Type I GRBs, which should in principle depend on the 
distributions of the masses of each member in the merging binary, 
the impact parameter, the collimation angle, and the viewing angle.
Taking other astrophysical objects (e.g. long or Type II GRBs) as
examples, we assume two general forms for the luminosity function
for Type I GRBs, i.e. a powerlaw (PL) 
\begin{equation}\label{PL}
\Phi(L)=\Phi_0\left(\frac{L}{L_b}\right)^{-\alpha},
\end{equation}
or a smoothed broken powerlaw (BPL)
\begin{equation}\label{BPL}
\Phi(L)=\Phi_0\left[\left(\frac{L}{L_b}\right)^{\alpha_1}
+\left(\frac{L}{L_b}\right)^{\alpha_2}\right]^{-1}~,
\end{equation}
where $\alpha$, $\alpha_1$, and $\alpha_2$ are the power law 
indices, $L_b$ the break luminosity, and $\Phi_0$ a normalization 
constant.  

\begin{figure}
\includegraphics[angle=0,scale=0.7]{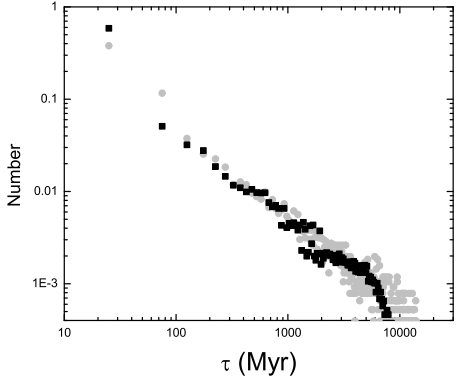}
\caption{A comparison of the simulated merger delay time distributions
between the standard population synthesis model (Belczynski et
al. 2008, grey) and the ``twin'' population synthesis model 
(Belczynski et al. 2007,
black).  Note the higher fraction of prompt mergers in the twin model.}
\label{belfit}
\end{figure}

Second, the $z$-distribution of Type I GRBs can be modeled
theoretically. 
For Type II GRBs, it is usually assumed that the volume
rate of GRBs follows the star forming history, since the delay
between the formation and death of a massive star is on the order of a
few million years, much shorter than the variations in the cosmic star
forming history or cosmological timescales.  For Type I bursts,
however, there is a delay, $\tau$, between star formation and the GRB.
In particular, for compact star merger scenarios, $\tau$ stands for
the delay between the creation of the binary
system, which follows the star forming history, and the eventual 
coalescence after the long decay
of the binary's orbit via gravitational radiation.  
So the redshift distribution of Type I GRBs can be determined by the 
star forming history distribution convolved by a distribution of the 
delay time scale $\tau$. This latter distribution is not
fully established theoretically, and we test the following four models 
that have been discussed in the literature:
\begin{itemize}
        \item Constant delay with dispersion:  
A $\delta$-function like delay with the center value 1-5 Gyr, and a 
normal dispersion of $\sigma = 0.3$ or $1.0$. These toy models are useful
to gain insight on how different delay time scales meet the constraints
of various models, but they are not very likely related to the true
delay distribution for compact star mergers.
        \item Logarithmic delay: Delays with a distribution
$P(\log(\tau))d\log(\tau) \sim$ const, which implies $P(\tau) 
\sim ~ 1/\tau$.  This empirical form was adopted by Piran (1992), 
Guetta and Piran (2006), and Nakar, Gal-Yam and Fox (2006). 
        \item Delay distribution from standard population synthesis: 
Belczynski et al. (2008) have modeled the NS-NS and NS-BH merger
delay time scales using their population synthesis code. We have
used their data and fit it with a $5^{th}$ order polynomial and
used this empirical model in our simulations. This model includes 
fast merger channels and allows for many short ($< 100$ Myr) 
mergers (see Fig.\ref{belfit}).  By applying this model,
it is assumed that the metallicity evolution effect 
does not play a significant role in defining the delay time
distribution of NS-NS and NS-BH mergers. This is evident by
comparing the calculated merger delay time distributions between 
solar metallicity (Belczynski et al. 2006) and $\sim 1\%$ solar
metallicity (Belczynski et al. 2008).
        \item Twin model for population synthesis: Belczynski et al. 
(2007) discussed another population synthesis model. 
This model incorporates the effect of twin binary systems (systems
with almost equal mass stars) and is characterized by an even larger
fraction of prompt mergers, with $\sim 70\%$ of the systems merging
within 100 Myr, as opposed to $\sim 40\%$ in the previous population
synthesis model. This most extreme prompt merger model would
set an upper limit on the rate of prompt compact star mergers
associated with star formation. We extract the delay
distribution data from Belczynski et al. (2007), and simulate
a distrbution of burst using Monte Carlo techniques (Fig.\ref{belfit}).
\end{itemize}
As will be evident later, most of the above models can interpret the 
short GRB data. We are then forced to consider the 
possibility that some or even most short GRBs are not Type I events 
but are instead related to massive stars (Type II). We therefore
consider the following two redshift distribution models as well. 
\begin{itemize}
        \item No delay (Type II): In this model, short GRBs
are assumed to follow star forming history of the universe, and are
therefore related to deaths of massive stars (Type II). We consider 
two variations on such a model. First, we leave the luminosity 
function of these short Type II GRBs as unknown, and constrain
it with the $L-z$ data. Second, we assume that these short
Type II GRBs share the same luminosity function as long Type II GRBs,
and use the established luminosity function of long Type II GRBs
(Liang et al. 2007; Virgili et al. 2009) to perform the $L-z$ and
$\log N-\log P$ tests. 
       
        \item Mixed Type I/II distribution: The observed short GRBs
        are a combination of a Type I population (with a delay
        distribution defined by the standard population synthesis
        model (Belczynski et al. 2008) or the twin model (Belczynski 
        et al. 2007) and a Type II population that follows
        the star forming history.  The fraction of bursts in each
        population is a free parameter and can be constrained from the
        data.
\end{itemize}
        
\begin{figure}
\plottwo{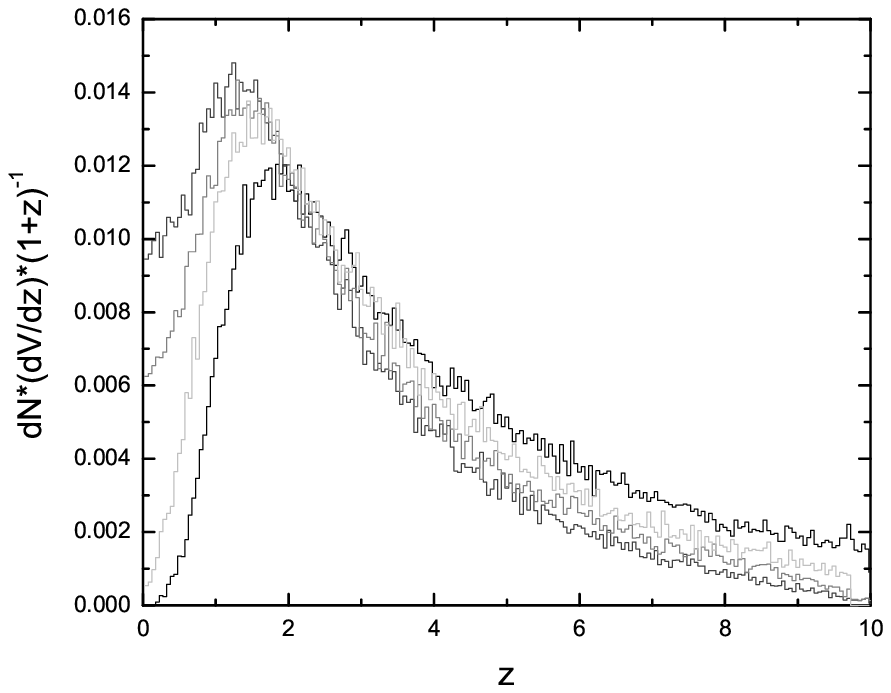}{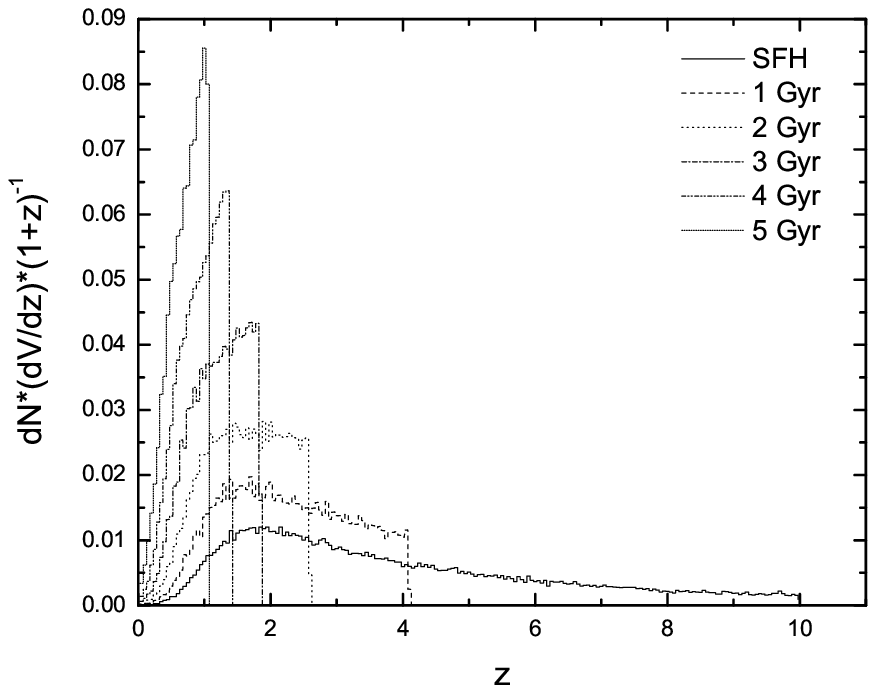}
\caption{Modified GRB redshift distributions (Eq.[1] integrated 
  over $L$) including the
  effects of cosmological time dilation and the comoving volume
  element, $dV/dz$. Different curves correspond to different models
  invoking different merger delay timescale distributions.  The left panel 
  shows a model that follows the star
  forming history (i.e. no merger time delay; black) as well as the
  population synthesis (standard, gray; twin, light
  gray) and logarithmic (dark gray) models.  The right panel shows various
  constant delay models as compared with the no delay model.
  All histograms contain the same number of bursts.}

\label{zdist}
\end{figure}

Once the value of the delay is assigned (in units of Myr) it needs to
be added to the previously simulated redshift to determine
the redshift of the GRB.
The redshift for the creation of the binary system,
$z_{creation}$ is assumed to follow the star forming history and is 
assigned from the SF2 model of Porciani and Madau (2001)
\begin{equation}\label{sf2}
R_{GRB}=23\rho_{\rm0}\frac{e^{3.4z}}{e^{3.4z}+22.0}.
\end{equation}
This redshift is then used to calculate the cosmological look-back
time by the following equation
\begin{equation}\label{lbtime}
{\int_{0}^{z_{creation}}t(z)dz} =
{\int_{0}^{z_{creation}} \frac{1}{H_{\rm 0}} \frac{1}{(1+z)
(\Omega_m(1+z)^3+\Omega_\Lambda)^{0.5}}}.
\end{equation}
With the values of this integral discretized over the simulated range
(z=0-10) in units of Myr, we then subtract the merger time delay and
re-convert to a redshift value with the same table.  Those bursts with
a negative lookback time (i.e. those that have not occurred yet) are
discarded.  The new redshift serves as the redshift of the merger and
the GRB, $z_{\rm GRB}$.  Figure \ref{zdist} shows how the redshift
distribution (including the co-moving volume element and cosmological 
time dilation terms) is affected by different models of the merger timescale
distribution.
      
With this formalism as a backdrop, we have enough information to
create a set of bursts, each one defined with a unique and random $(L, z)$
pair, which can then be passed through a series of filters that mimic
a detector and then compared to the observed distribution.  These
simulations are similar to those conducted in Virgili et al. (2009),
and here we summarize the most significant points.
        
Fundamentally, Monte Carlo simulations rely on random numbers, and
although it is possible to create true random numbers with a device that
uses a stochastic process (e.g. thermal noise), pseudo-random numbers
are much more convenient in terms of ease of use and possibility for
exact repetition of simulations.  In this code we utilize the
SIMD-oriented Fast Merssene Twister, created by Mutsuo Saito and
Makoto Matsumoto \cite{sandm08} of Hiroshima University.  It is
specifically designed for use with scientific Monte Carlo simulations,
producing long strings of random numbers with a period of anywhere
from $2^{607}-1$ to $2^{216091}-1$.
        
In order to compare the simulated output with observations, it is
necessary to be in the same band as the detector with which one is
observing (i.e. the $k$-correction).  In order to achieve this, we
assume every burst has a Band Function spectrum \cite{band93}, with
spectral indices $\alpha = -1.0$ and $\beta= -2.3$ below and above a
characteristic energy $E_0$.  Although some BATSE and Swift bursts 
have been adequately fit with an exponential cut-off power law 
(Ghirlanda et al. 2004; Sakamoto et al. 2008), this is very likely 
because the flux above $E_p$ is too low to constrain the high energy
photon index of the Band function. 
Recent Fermi observations suggest that for bright enough bursts,
the Band function can fit the spectrum for both long 
(e.g. 080916C, Abdo et al. 2009a) and short bursts (e.g. GRB 090510, 
Abdo et al. 2009b).  In the case of GRB 080916C, the Band spectrum 
extends several orders of magnitude in energy
and supports our choice of the intrinsic spectrum for the simulated
bursts.  The characteristic energy correlates with the peak of the
$\nu F(\nu)$ spectrum by the relation $E_p=E_0(2+\alpha)$, 
which we assign from the relation proposed by Liang et el (2004)
\footnote{Although
this correlation was derived for long GRBs, it has been found that
short GRBs share the similar $E_p - L$ correlation 
\cite{ghirlanda09,zhang09}. The  
$E_p - E_{\gamma,iso}$ correlations for the two categories are 
very different, mainly due to the shorter durations of short GRBs 
with respect to long GRBs.}
\begin{equation}\label{EpLiso}
 E_p/200 {\rm keV}=C^{-1} (L/10^{52} {\rm erg\ s}^{-1})^{1/2}
\end{equation}
where $C$ is randomly distributed in [0.1,1].  This energy can then be
used for the $k$-correction from the simulated bolometric luminosity
in the rest-frame $1-10^4$ keV band (based on a certain luminosity
function) into an arbitrary detector bandpass spanning the energy
range ($e_1$, $e_2$). The $k$-correction parameter is defined by
\begin{equation}\label{kcorr}
k=\frac{\int_{1/(1+z)}^{10^4/(1+z)}EN(E)dE}{\int_{e_1}^{e_2}EN(E)dE}.
\end{equation}

The last step is to incorporate the detector threshold condition.  
 For Swift, similar to Virgili et al. (2009), we
apply the fluence threshold (Sakamoto et al. 2007)
\begin{equation}
F_{th} \sim (5.3\times 10^{-9}~{\rm erg~cm^{-2}~s^{-1}})T_{90}^{-0.5}~
\label{threshold}
\end{equation}
in the 15-150 keV band as an approximation to the Swift trigger.  This is because any valid rate trigger requires a statistically significant excess both in the rate and
in the fluence domain, the latter being particularly
stringent for short duration bursts. To
calculate fluence from luminosity, we have randomly assigned a 
duration $T_{90}^{\rm short}=0.33 \pm 0.21$s based on the BATSE
sample statistics (Kouveliotou et al. 1993). 

Another test is the peak-flux distribution, i.e. $\log N - \log P$.
Since redshift information is not needed, the sample is much larger. 
We consider both the Swift and BATSE short GRB samples. The latter is included because short GRBs were originally defined 
using the BATSE sample and if one wants to make the claim that 
``short GRBs are produced
from compact star mergers'', one should make the case that both the
BATSE and Swift short GRB samples are consistent with this hypothesis.  For both distributions we filter the observational sample by utilizing a truncation in photon flux which is determined by the detector efficiency, as detailed in Loredo \& Wasserman (1998), in order to provide an accurate statistical subsample of bursts.  Both samples, coincidentally, have a cutoff of approximately 1 $\rm ph~cm^{-2}~s^{-1}$ above which the detector is sensitive, in the 50-300 keV and 1-1000 keV bands for BATSE and BAT, respectively (for BATSE see Loredo \& Wasserman, for BAT see Band 2006, fig 3b).  We apply this simple rate trigger to our simulated sample in order to compare similar subsamples.  For the Swift sample, this is essentially consistent with the more complicated fluence trigger criterion discussed above for the $L-z$ constraints and we adopt this rate trigger criterion for the $\log N - \log P$ analysis for the sake of simplicity.

\section{Results}

The set of short bursts with known redshift is, to date, relatively
small.  We collect all the short GRBs with $z$ information
up to May 2009. The sample is compiled in Table 1. 
There are additional short bursts that have redshift claims (e.g. GRBs 000607, 
051210, 060313, 060502B, 061201, 070809, and 080503), but we do not
include them either because the redshift is uncertain, or because the 
burst is too faint to extract good spectral parameters so that no 
reliable luminosity can be derived. 
For the sample we present, we assume
that the redshift values are all correct, but caution
about the small chance of mis-identification due to afterglow/host
chance coincidence \cite{cobb08}. Among the highest redshift GRBs in 
this sample, GRB 070714B has $z=0.923$ (Graham et al. 2009), and
GRB 090426 has $z=2.6$ (Levesque et al. 2009). GRB 060121 has two
uncertain redshifts $z=1.7, 4.6$ (de Ugarte Postigo et al. 2006), 
and we take the smaller value $z=1.7$. Some studies (e.g. 
Berger et al. 2007; Berger 2009) suggest that there are more 
short bursts at these high redshifts. This would further strengthen the
argument presented in this paper, namely, 
most short GRBs track the star formation 
history of the universe.

\begin{figure}
\plotone{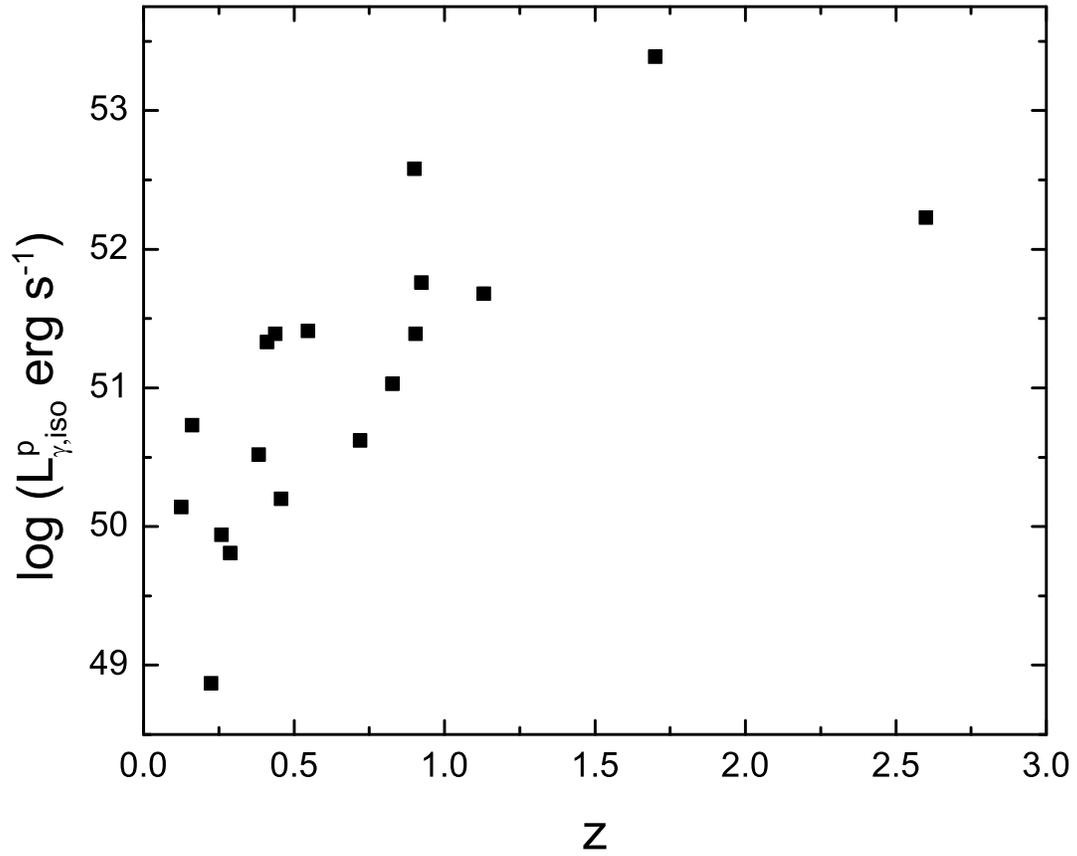}
\caption{Sample of the $z$-known short-hard GRBs detected in the Swift
  era.  The redshifts are plotted
  against peak isotropic gamma-ray energy, $L$. This distribution is
  used to constrain luminosity function of various redshift
  distribution models.}
\label{obs}
\end{figure}

\begin{figure}
\includegraphics[angle=0,scale=0.7]{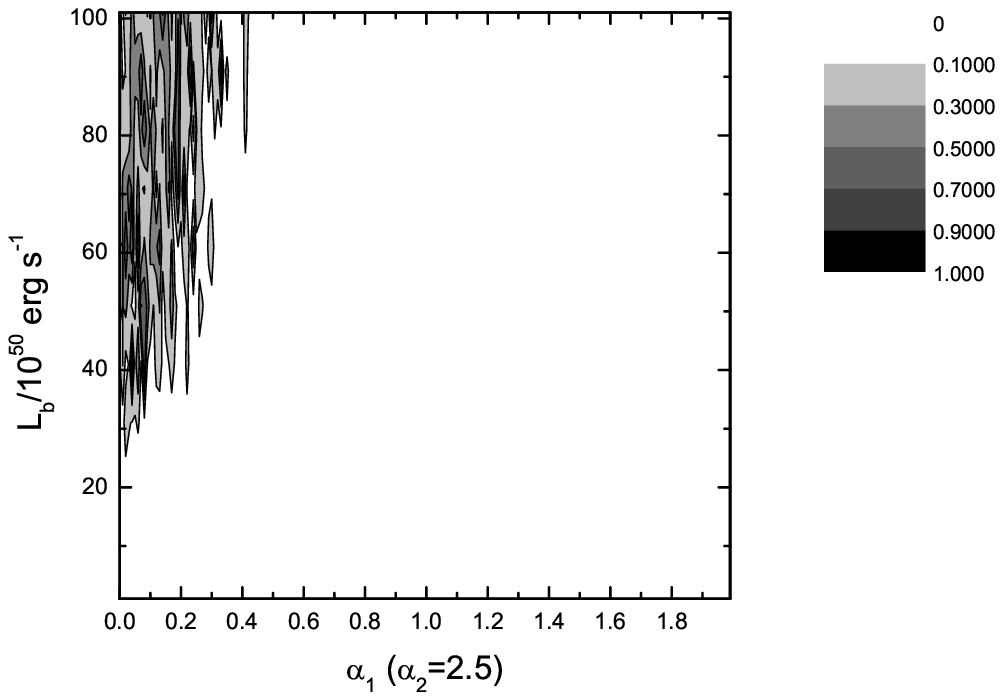}
\includegraphics[angle=0,scale=0.7]{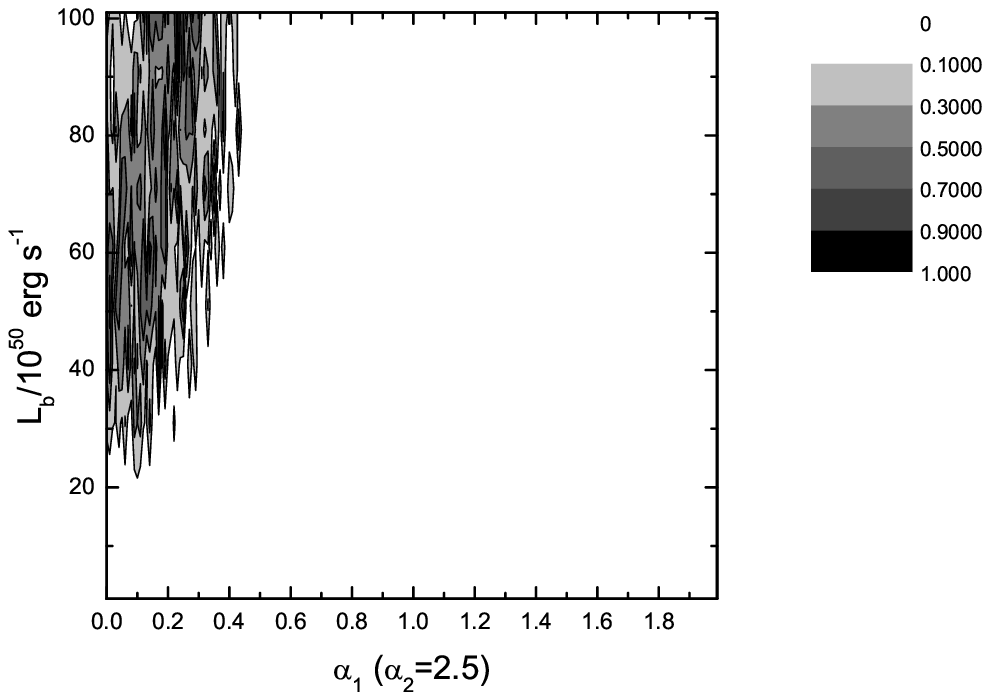}
\includegraphics[angle=0,scale=0.7]{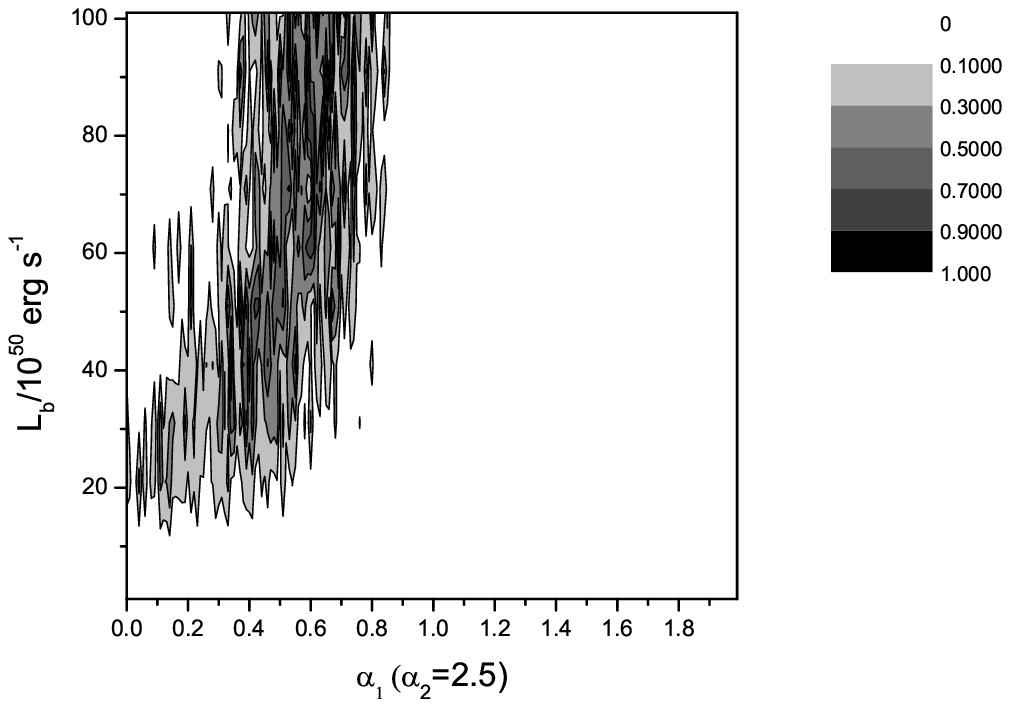}
\includegraphics[angle=0,scale=0.7]{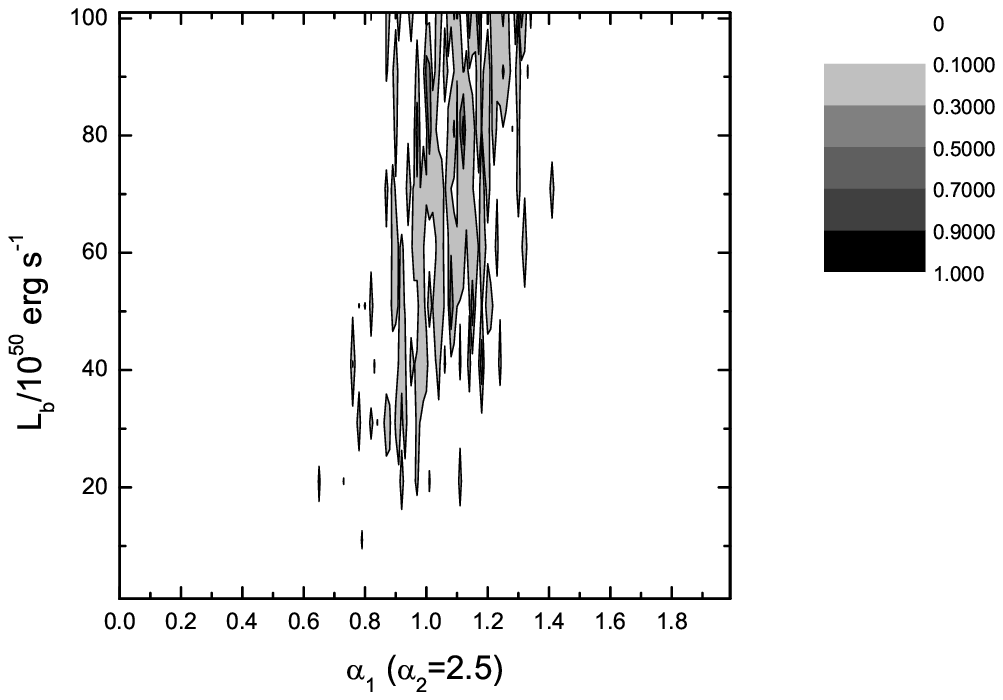}
\caption{A series of contours displaying the total KS probability,
  $P_{KS,t}$ of varying luminosity function parameters (break
  luminosity, $L_b$ and pre-break power-law slope $\alpha_1$) 
  derived from the $L-z$ constraints for a sample of redshift
  distribution models.
  (a) the logarithmic model, (b) the standard population
  synthesis model, (c) the ``twin'' population synthesis model, 
  and (d) the no delay model.  
  Darker indicates higher KS probabilities for consistency with the 
  observed $L-z$ distribution.}
\label{contours}
\end{figure}

Utilizing the models presented in the previous section, we ran various
sets of simulations combining different luminosity functions and
time delay distributions.  Varying the luminosity function
parameters for a particular time delay distribution, one can
compare the model predictions of GRBs in the $L-z$ space
with the $L-z$ distributions of the data (see Fig.\ref{obs}).
Each comparison utilizes the 1D KS probability in $z$ ($P_{KS,z}$), 
1D KS probability in $L$ ($P_{KS,L}$), and the total KS probability, 
defined as $P_{KS,t}=P_{KS,z}\times P_{KS,L}.$
For a single power law
LF (Eq.[3]), varying $\alpha$ leads to a distribution of $P_{KS,t}$ 
and the maximum $P_{KS,t}$ defines the most likely $\alpha$ value. 
Since the observed short GRBs seem to have an upper cutoff in the $L$
distribution, it is more reasonable to adopt a smoothed broken
power law LF (Eq.[4]). However, since there are three free parameters
($\alpha_1$, $\alpha_2$, and $L_b$), it is difficult to constrain
all three parameters. We therefore fix $\alpha_2 = 2.5$,
and constrain $\alpha_1$ and $L_b$ together using the $L-z$
criterion\footnote{The reason of fixing $\alpha_2$ is because the other 
luminosity function parameters ($\alpha_1$ and $L_b$) have a greater 
effect on the simulated burst luminosity distributions (since the
number of higher-$L$ GRBs is much smaller than that of lower-$L$
GRBs). We again follow the arguments in Liang et al. (2007) and 
Virgili et al. (2009) as guides in choosing a constant $\alpha_2$ 
in the range of 2-2.5.}.
This results in a series of $P_{KS,t}$ contours in the $(\alpha_1, L_b)$
plane, an example of which is shown in Fig.\ref{contours}, from which we can 
infer the best fit parameters for that particular model. 
Using these parameters, we can then construct a simulated 2D
$L-z$ graph with both the observed and simulated data plotted.

Our next constraint for the simulated bursts is consistency with 
the BATSE and Swift $\log N - \log P$ distributions.  The BATSE 4B
catalog \cite{paciesas99} has 309 short bursts  with 
$T_{90} < 2 s$ (on a 64 ms timescale).  When binned, the distribution 
gives a power law of slope of -1.12, extending from about 1-50 
$\rm ph~cm^{-2}~s^{-1}$ (in the 50-300 keV band), disregarding the 
turnover at low photon flux, which is likely an artifact of the 
detector.
Adopting the 1 $\rm ph~cm^{-2}~s^{-1}$ threshold as discussed
above, we get a reduced BATSE short GRB sample with 271 bursts.  
This is the first sample we use to compare against simulation.
The second sample is the Swift short GRB sample above the
1 $\rm ph~cm^{-2}~s^{-1}$ count rate threshold. We get 31 bursts
in this sample. The third sample includes all $z$-known Swift short
GRBs above the 1 $\rm ph~cm^{-2}~s^{-1}$ count rate threshold.
This sample only has 12 GRBs.

 For any $z$-distribution model, we adopt the most probable
luminosity function derived by the $L-z$ constraint and simulate the
$\log N - \log P$ distribution in the various detector bands. The
simulated photon flux output is screened by a simple cut at the BATSE
and Swift threshold of 1 $\rm ph~cm^{-2}~s^{-1}$. The model results
are compared with the observed data by the k-sample Anderson-Darling
(AD) test \cite{sands}, which is more reliable than the KS or $\chi^2$
tests when distributions have smaller numbers and/or might be drawing
from the tails of the unknown underlying distribution. This test is
similar to the KS test in that the null hypothesis is that the
distributions come from the same underlying yet unknown distribution,
but uses a more reliable test statistic.  In order to accept the null
hypothesis at the 95\% confidence level, the statistic (which we call
the T-statistic, see Table 2) must lie below a value of 1.96.  In
addition, the AD test is a test on the un-binned photon flux
distribution, which is always more statistically reliable than the
binned distribution.  The figures are binned in log-log space for easy
inspection and for consistency with the literature.  The above
procedure allows for a self-consistent check of every model through
the joint $L-z$ and $\log N - \log P$ analysis.  Any correct
population model (including $z$ distribution and luminosity function)
should be able to show consistency in all data sets ($L-z$
distribution for the $z$-known short GRB sample, and the $\log N-\log
P$ distribution of the BATSE and Swift short GRB samples).

Next we break down the results by merger time delay model and comment 
on the constraints imposed by the observations. Table 2 is a 
comprehensive list of test statistics and P-values for the relevant 
models in this analysis.

\subsection{Constant merger time delay}

We start with the constant merger delay models. 
The five models we considered are $\tau = 1, 2, 3, 4, 5$ Gyr, each 
with a gaussian scatter of 0.3 Gyr or 1.0 Gyr around these values. 
These models are not realistic compact star merger models.
Most NS-NS systems in our galaxy have a predicted merger delay 
time scale longer than the Hubble time scale. Any realistic
NS-NS merger delay time distribution model should include
these systems. In any case, these models include a long time 
delay tail extending to Hubble time, and cannot be modeled by
the constant delay models as discussed in this section. 
By studying these models,
one would be able to diagnose how mergers at different delay time 
scales contribute to the global $L-z$ and $\log N - \log P$
distributions, so that the results invoking more complicated 
models (e.g. those invoking population synthesis) can be better
understood.

Except for the 1-2 Gyr models, all other models demand a very shallow
$\alpha_1$ to account for the observed $L-z$ distribution.
When combined with the corresponding $z$-distribution, this
always leads to a very shallow BATSE $\log N - \log P$ that is inconsistent
with the data. The reason is that the merger models give a clustering 
of bursts at low-$z$ (Fig.2) so that the shape of the luminosity
function carries significant weight in defining the shape of 
$\log N - \log P$. This is in contrast to the case of Type II GRBs, 
which are spread in a wide range of $z$ so that the shape of 
luminosity function play a less important role in defining the shape 
of $\log N - \log P$ 
\footnote{In an
extreme, for a pure Euclidean geometry, the $\log N - \log P$ always
has a slope of -3/2 regardless of the shape of luminosity function.}.   
The BATSE constraints show only the 2 Gyr model as passing the 
AD test.  The constraints from the Swift sample, however, are much 
more forgiving for these models, allowing for 1-4 Gyr delays as 
acceptable fits. The results are insensitive to the assumed 
Gaussian scatter (0.3 Gyr vs. 1.0 Gyr).  A breakdown of the various tests is illustrated in Fig. \ref{2gyr}.
By combining all the contraints, we conclude that only the
2 Gyr constant merger time delay model is consistent with
the data. 

\subsection{Logarithmic and Population synthesis}

More realistic compact star merger models are those that 
introduce a distribution 
of the merger delay time scales. We discuss the logarithmic
distribution model and two more detailed models involving population
synthesis \cite{bel08,bel07}. These models have the advantage to
allow for a range of merger times that can lead to 
bursts in diverse host galaxy types. 

The logarithmic and standard population synthesis models affect the redshift 
distribution in similar ways and have similar results in all tests, 
and are therefore discussed together. The constraints from the 
observed $L-z$ distributions all imply a very shallow 
luminosity function for these two models, 
generally having a slope of -0.2 or even larger, 
with moderate consistency (40-50\%) for a broken power law luminosity 
function (see Fig.\ref{contours}).  The reason for a shallow luminosity
function is to avoid overproducing nearby low-$L$ short GRBs.
These shallow luminosity functions severely 
overproduce at high photon fluxes, giving a much shallower 
$\log N - \log P$ curve than observed. These models are therefore
not favored by either the BATSE or Swift short GRB data.

The special population synthesis ``twin model'' as discussed above
allows for an even larger fraction
($\sim 70\%$) of prompt mergers as compared to the standard population
synthesis model  ($\sim 40\%$).  This affects the observed burst 
distribution by removing many of the higher merger timescale bursts 
and creating a distribution closer to one with no delay from the star 
forming history. This model has good consistency with the observed $L-z$
distributions and requires a steeper luminosity function slope than
the previously discussed models. The $\log N-\log P$ is 
steeper as well, but not sufficiently steep for consistency with 
the BATSE sample, although consistency is achieved for the
smaller Swift sample. Taking both constraints together, this model 
alone cannot adequately reproduce the observations.

\subsection{No delay (Type II)}

The above analysis leads us to conclude that the hypothesis that
``all short GRBs detected by BATSE and Swift are of the compact
star merger origin'' is not justified, and that one needs 
to seriously explore alternative models. We first explore
the other extreme, namely, that all short GRBs track the star forming 
history of the universe. Similar to long GRBs, they are related
to deaths of massive stars (Type II). Such a hypothesis is already
disfavored by the host galaxy data of some short GRBs (e.g.
Fong et al. 2010). We test this model mainly to see how close
the real short GRB $z$-distribution is to the star formation
history. We discuss two variations of this model. 
The first approach is 
to leave the luminosity function of these short Type II GRBs
as unknown, and constrain it from the $L-z$ data. This 
approach is similar to the previous discussion of the other delay models.
The implied slope of the luminosity function is, as expected from 
the general trend of steepening with decreasing delay time, steeper 
than the lowest constant $\tau$ model, weighing in at $\alpha_1 = 
1.42$.  The consistency with the observed $L-z$ distributions is 
low, about 20\%, but not sufficient to completely rule out an 
association.  The implied $\log N - \log P$ is inconsistent 
with observations for both the BATSE and the Swift samples 
(Fig \ref{lnlp}a).

In the second approach, we also consider the possibility that
these short Type II GRBs have a same parent population as the
known long Type II GRBs, so that we can use the established
luminosity function of long Type II GRBs (e.g. Liang et al.
2007; Virgili et al. 2009). As expected, this model is securely 
ruled out by the $L-z$ constraints although the $\log N-\log P$ 
distribution is consistent with the observations 
(see Fig. \ref{lnlp}c).

In summary, attributing all
short GRBs to Type II GRBs is not justified. This is expected 
since several Type I Gold Sample GRBs have been identified
(Zhang et al. 2009), suggesting that at least some short GRBs
should be of the Type I origin.

\subsection{Mixed population model}

Since the realistic compact star merger models detailed in the 
previous section are unable to 
reproduce all observational tests, we are forced to consider 
the possibility that the observed short
GRBs include both compact star merger (Type I) events and events
that are associated with massive stars (Type II). 
Although this possibility looks ad hoc at first sight, it may be 
already implied by the data. Zhang et al. (2009) discussed
various criteria to assign a progenitor to a GRB and 
concluded that the often used $T_{90}$ is 
not necessarily an informative parameter 
to define the physical category of a GRB. After discussing a series
of multiple observational criteria, they applied the criteria that
most directly carries the progenitor information and defined a Type I
Gold Sample. They found that the Gold Sample bursts are relatively
long (and most have extended emission), not particularly hard
and that they are not a fair representation of 
the short hard GRB sample. On the other hand, none of the high
luminosity short/hard GRBs have been
found in elliptical or early type galaxies. Zhang et al. (2009) 
therefore speculated
that some or even most high-luminosity short GRBs are of the Type 
II origin. Our above analysis strengthens that conclusion and
demands a more serious investigation of the mixed population in
short GRBs. 

We test this hypothesis by analyzing various merger timescale
distributions that are a combination of Type I and Type II bursts.
We consider a mix of Type I GRBs and the ``classical'' 
Type II GRBs, namely,
those same Type II GRBs that account for the long-duration
GRBs\footnote{ In principle, the short Type II GRBs can be
different from long Type II GRBs by having a different type
of massive star progenitors. This would give too many unknown 
parameters to be constrained by the data.}.
In this approach,
the Type I bursts follow the distribution of merger delay time scales
as predicted by the population synthesis models of Belczynski et
al. (2008, 2007), and the Type II component tracks the star forming 
history of the universe and obey the known luminosity function of the 
long Type II population, as discussed in Virgili et al. (2009) and 
Liang et al. (2007). The logarithmic distribution is very similar 
to the standard population-synthesis derived distributions
(Belczynski et al. 2008), so we do not discuss such an option
explicitly.

We begin with the limiting case of 100\% Type II GRBs.  As 
mentioned in \S 3.3, this model (the ``no delay'' Type II model)
shows good consistency with the observed
$\log N-\log P$ distribution (Fig. \ref{lnlp}c)\footnote{There is 
a small excess at the high
photon flux end. In view of the log scale involved, this only gives
an excess of bursts above $100\rm~ph~cm^{-2}~s^{-1}$ that is below
2.}, but is securely ruled out by the $L-z$ constraints.

Next we test how different amounts of mixing can effect
the distribution of bursts.  To achieve this goal we do a series of
simulations with different amount of mixing between the population
synthesis Type I GRBs (based on the standard model of Belczynski 
et al. 2008) and the classical Type II bursts.  A `20\% mix', for example, indicates 20\% Type II and
80\% Type I.  The consistency with the $L-z$ constraints peaks
around a 75\% mix then falls off rapidly (with a few patches in the
significance contours) near 90\%.  The
corresponding $\log N-\log P$ distributions are plotted in 
Fig.\ref{lnlp}c
together with the limiting case model composed entirely of 
classical Type II bursts. As shown, all
models are too shallow or have hidden inconsistencies that are 
picked up with the AD test to be consistent with observations for 
the BATSE sample.  This is expected in view of 
the lack of consistency with the un-mixed distribution.  For the 
smaller Swift sample, the low mix models do equally as badly, but 
they begin to show consistency above 75\%.

The ``twin'' population synthesis model (Belczynski et al. 2007) 
predicts an even higher
fraction of prompt mergers ($\sim 70\%$ with $\tau<100$ Myr), causing
the redshift distribution to differ less from the one of pure Type II
bursts. This is likely to affect the amount of mixing and the
steepness of the $\log N-\log P$. We continue our analysis with the 
same procedure as the standard population synthesis model, slowly
increasing the amount of mixing with the Type II bursts and observing how
the results compare with the observations.  As expected, the increase in
prompt mergers decreased the amount of mixing needed for a good fit
to the $L-z$ data. The peak probability is achieved for 20-30\%
Type II mixing, and a Type II mixing higher than 60\% is securely
ruled out by the $L-z$ data. The $\log N-\log P$
distribution is still too shallow up to a 30\% mix to be consistent
with the BATSE sample (Fig \ref{lnlp}d). 
The Swift constraints are once again more forgiving, showing 
consistency with the 10-40\% mix models.  Taken together, we find 
consistency with the observed distributions in the 30-40\% range
for Type II mixing.  Figure \ref{mix30twin} shows a graphical breakdown of the tests for a 30\% Type II mix model.

\section{Conclusions and Discussion}

\begin{figure}
\plottwo{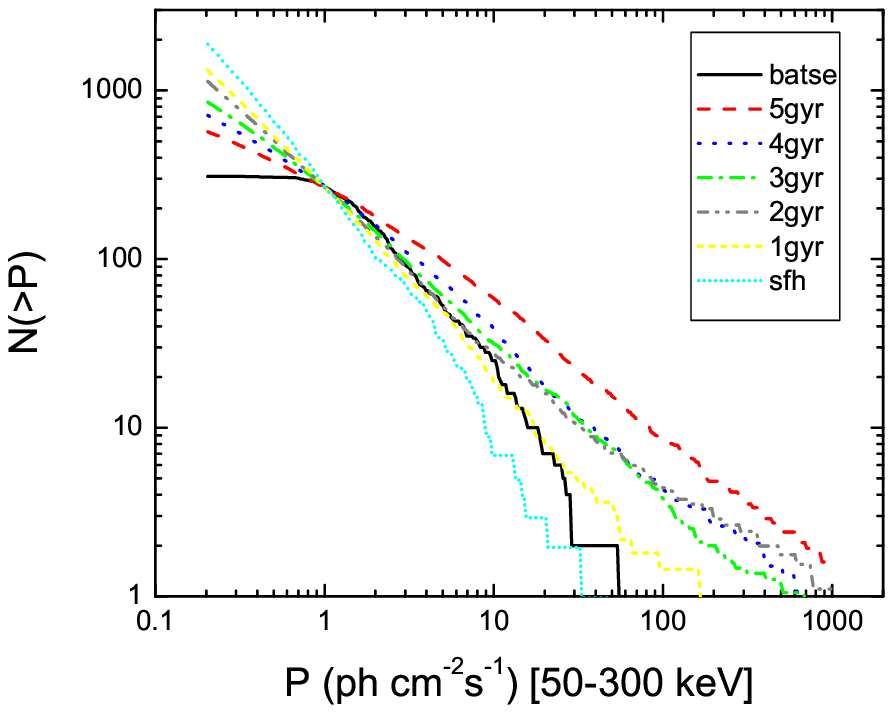}{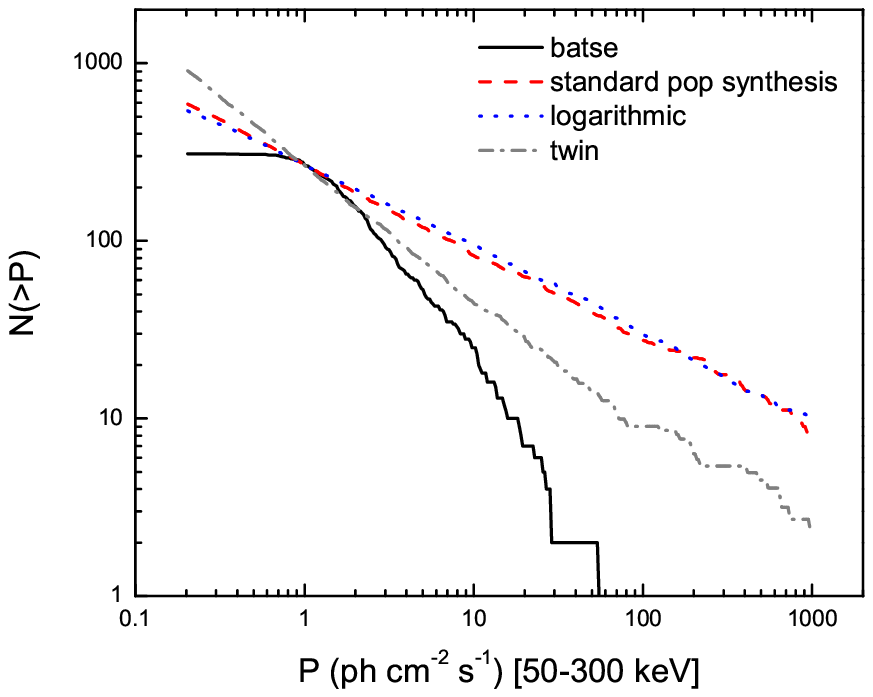}
\plottwo{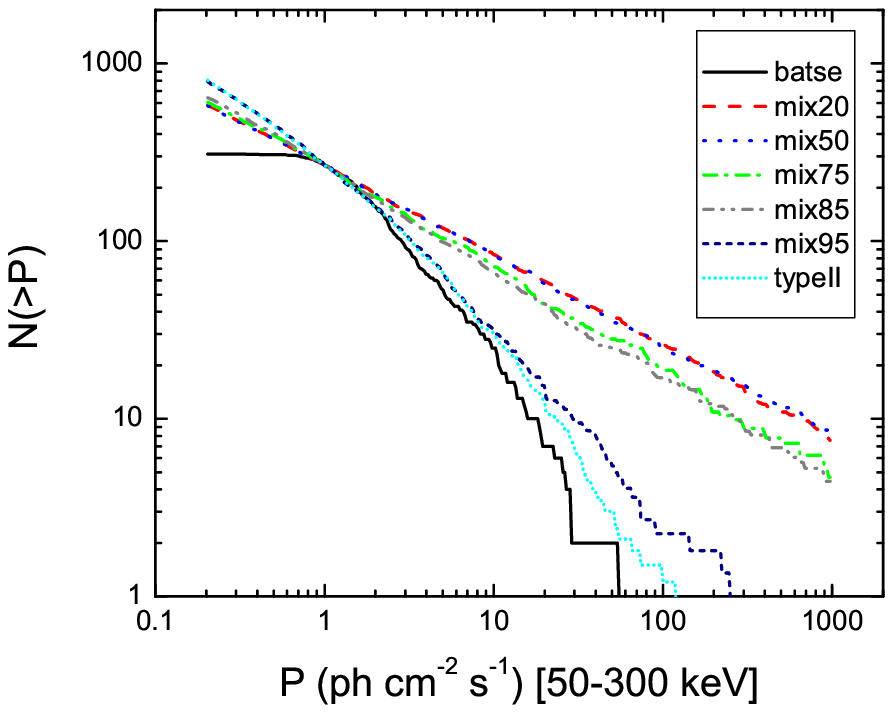}{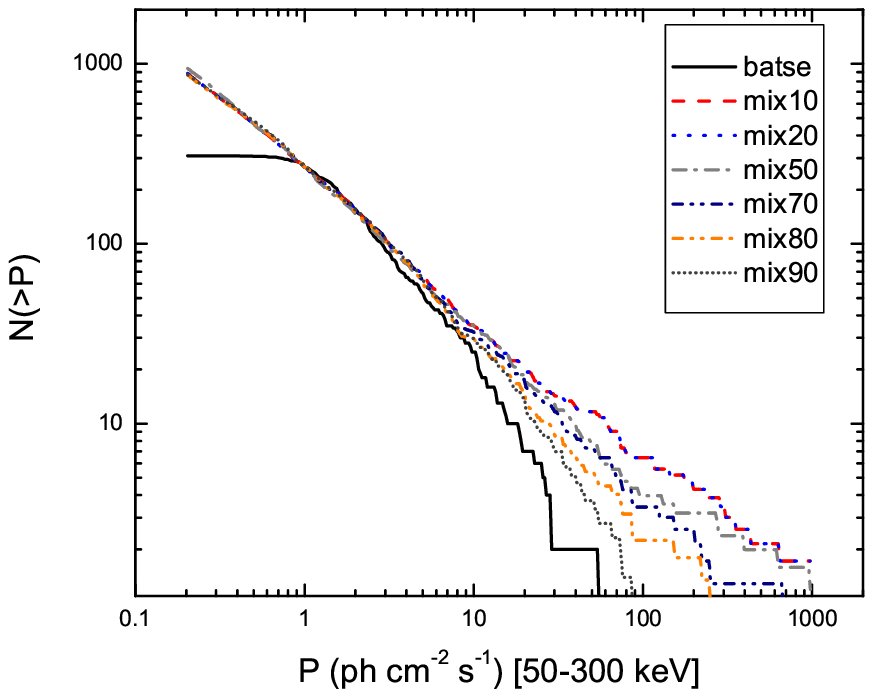}
\caption{Comparison of the $\log N-\log P$ distributions for the
  various models with the observed BATSE curve.  
  (a) various constant delay merger models ($\sigma=0.3$ are shown.  
  Curves for $\sigma=1.0$ are similar and therefore not included 
  in the figure); (b)
  the standard population synthesis, logarithmic and twin models.  
  (c) mixed models with classical Type II's (with long Type II 
  luminosity function) and Type I's with standard population
  synthesis time delay distribution; (d) mixed models with 
  classical Type II's and Type I's with time delay distribution
  predicted by the ``twin'' population synthesis model. 
  The notation ``mix20'' standards
  for 20\% Type II (and 80\% Type I) for both panels (c) and (d).   
  Few models pass the BATSE constraints, with the exception of: 
  (1) the 2 Gyr model (both $\sigma=0.3$ and $1.0$); and (2) the 
  30\% and 40\% Type II-twin mix models.  See Table 2 for test 
  statistics and P-values for various models. } 
\label{lnlp}
\end{figure}

With the extensive afterglow follow up observations of short GRBs
in the Swift era, the sample of $z$-known GRBs has significantly expanded.
This sample, together with the BATSE short GRB sample, can be used to
constrain the luminosity function and redshift distribution of short
GRBs by means of reproducing the observed $L-z$ distribution and
$\log N - \log P$ distribution. In this paper, we have performed a 
detailed analysis on a list of models using Monte Carlo simulations. 
Our results can be summarized as follows:
\begin{itemize}
\item The hypothesis that ``all short hard GRBs are of the compact 
star merger origin'' is disfavored by the data. In particular,
the merger time delay models derived from population synthesis 
(Belczynski et al. 2006, 2008, 2007) or using some empirical 
(e.g. logarithmic) formulae
all demand a very shallow luminosity function in order to
satisfy the $L-z$ constraint. This is because a steeper luminosity
function would over-produce low-$z$, low-$L$ short GRBs that are
not observed. Such a required shallow luminosity function, combined 
with the redshift distribution derived from the merger delay time
distribution, leads to a very shallow predicted $\log N - \log P$,
which is  inconsistent with the BATSE $\log N - \log P$
data. 

\item Among constant delay models, those
with large delays (3-5 Gyr) suffer the same problem. Only the
2 Gyr model can satisfy both the $L-z$ and the $\log N-\log P$ 
(for both the BATSE and Swift samples, see Figs. \ref{lnlp} and \ref{2gyr}) constraints. Such a model
is however not a realistic model for compact star mergers,
since most NS-NS binaries in our galaxies are found to have
a merger delay time scale longer than Hubble time, not 
narrowly clustered near 2 Gyr.
\item A model that invokes no merger delay (Type II) 
can better satisfy both the $L-z$ and $\log N - \log P$ constrains. 
However, a full consistency cannot be achieved. For a model
invoking a Type II population with free luminosity function,
the best fit luminosity function from the $L-z$ constraint
predicts too steep a $\log N - \log P$ as compared with data.
A model that invokes a luminosity function of the classical
long Type II GRBs is securely ruled out, since it cannot 
reproduce the observed $L-z$ distribution.
\item After considering the various tests detailed above, it seems 
that if one insists on the compact star merger model for (some) short 
GRBs, the data then demand a significant mixing of Type I and Type 
II GRBs in the observed short GRB population. This is in consistent 
with the argument presented in Zhang et al. (2009) who argued that
the Type I Gold Sample is not a fair representation of the 
BATSE short/hard GRBs. For models mixing Type I bursts 
predicted from the standard population synthesis model and classical
Type II bursts, no model can pass the BATSE $\log N - \log P$ test, 
and only models between 75\% and 90\% Type II mixing show consistency 
with the Swift $\log N - \log P$ and the $L-z$ constraints.  For the
most extreme ``twin'' population synthesis model invoking a much
larger fraction of prompt mergers, consistency in all tests is
achieved with a Type II mix of 30-40\%. (See Figs. \ref{lnlp} and \ref{mix30twin})
\end{itemize}

The above analysis indicates that there are
two solutions for the short GRB $z$-distribution to satisfy the
$L-z$ and $\log N - \log P$ constraints. The first one is $\sim 2$ Gyr 
constant delay distribution with some scatter. The second one is to 
have a wide range of delay time scales with respect to star formation,
but the distribution is heavily tilted towards short delays, namely,
over 80\% short GRBs should have a delay time scale shorter than 
$100$ Myr. This can be translated into two possible scenarios regarding
the progenitor of short GRBs.

(1) If one takes the widely discussed compact star merger model 
for short GRBs, then this model alone cannot account for all
short GRBs. This is because the compact star merger time scales
cannot be clustered around 2 Gyr (in view of the galactic NS-NS
population), and because any reasonable merger delay time scale
distribution cannot give the high percentage ($>80\%$) of prompt
mergers. The standard population synthesis model only has $\sim 40\%$
such prompt mergers. The most extreme ``twin'' population synthesis
model has a prompt merger fraction $\sim 70\%$, still not enough
to satisfy the data constraints. Inevitably, one has to consider
a superposition picture, namely, besides these merger-origin Type
I GRBs, there are Type II GRB contamination in the short GRB
population. Our analysis suggest that for the ``twin'' model,
one still needs $30\% - 40\%$ Type II GRB contamination in the
merger-origin Type I GRBs. The fraction of prompt mergers is
therefore $(30\%+70\% \times 70\%) - (40\% + 60\% \times
70\%)  \simeq (79\% - 82\%)$ of events with delay time scale 
$< 100$ Myr. This is consistent with the general $80\%$
constraint. We note Cui et al. (2010) obtained a similar
contamination fraction based on a different criterion
(afterglow location in the host galaxy).

 (2) Alternatively, the entire short GRB population may be related
to a different type of progenitor with a typical delay time scale
$\sim 2$ Gyr  with respect to star formation history. Such a 
progenitor can be still of the Type I origin (explosions from 
compact stars), but is not from compact star mergers. One 
possible candidate may be accretion induced collapse of NS
in binary systems \cite{qin98,dermer06}.

Our analysis is based on $L$, $z$, and $\log N-\log P$ distributions
of short GRB samples (both Swift and BATSE). There are other 
observational facts of short GRBs that are not considered in our 
modeling. Nonetheless, below we comment on how our results may be
compatible with those observations. 

\begin{itemize}
\item A small fraction of short GRBs have elliptical/early type 
host galaxies, the two best cases are GRB 050509B (Gehrels et al. 
2005; Bloom et al. 2006) and GRB 050724 (Barthelmy et al. 2005; 
Berger et al. 2005). This is consistent with our suggestion that 
most short GRBs have short delay with respect to star formation.
\item Fong et al. (2010) found that the relative location 
distribution of short GRBs in their host galaxies is different 
from that of long GRBs. Instead of tracking the brightest regions
of the host galaxies, they under-represent the light distribution
of their hosts. Their sample includes 10 GRBs, which are mostly 
nearby low-luminosity short GRBs. According to our first scenario
(superposition scenario), the 
high-luminosity, high-redshift short GRBs are more likely 
Type II GRBs, but their host galaxies are not well studied.
Even within the Fong et al. sample, most hosts are
late-type galaxies. The difference with long GRBs in host
location distribution only states that their progenitors are
different from the long GRB progenitors. The short Type II GRBs
can be related to other types of massive stars, which are
different (e.g. in mass, spin, and/or metallicity) from the
the progenitors of long Type II GRBs. Alternatively, according
to our second scenario, all short GRBs may belong to a different
type of Type I GRBs.
\item Recent work by \cite{landb} suggest that delay times 
for short-hard bursts as derived from IR and optical observations 
of 19 bursts seem to imply different time delays for bursts occurring 
late- and early-type galaxies with median delays of $\sim$0.2 and 
$\sim$3 Gyr, respectively.  The latter estimate, as well as an 
estimate for all galaxy types, falls within the limits of our 
2 Gyr constant plus scatter models especially if one considers a 
large scatter around the median.  
It would be also consistent with our first scenario, namely,
the observed population is a mix of a significant fraction of
prompt mergers along with some mix with Type II GRBs, along with
a tail of long-delay mergers.
\item Deep upper limits for SN association have been established
in a few short GRBs. These are nearby low-luminosity short
GRBs, which we also expect that they are of the Type I origin.
So far there is no constraint on the SN association for high-$L$
short GRBs. Some of these GRBs can be short Type II candidates 
according to our first scenario (superposition). According to
our second scenario (2 Gy delay), all short GRBs are not expected
to be accompanied by SNe.
\item The metallicity of short GRB hosts is systematically different
from that of long GRBs. Again the sample is mostly for nearby
low-luminosity short GRBs. More data for high-$z$, high-$L$ GRBs
would help to distinguish between the two scenarios discussed in
this paper.

\end{itemize}

 Our results imply that if (some) short GRBs are from compact
star mergers, the merger rate that give 
rise to short GRBs is smaller by $30\%-40\%$
than previously estimated based on the assumption that all short
GRBs are due to compact star mergers \citep[e.g.][]{ngf06}. If
these merger events are also gravitational wave bursts, then
the rate of gravitational wave bursts that are associated with 
GRBs is also lower by the same fraction.  Alternatively, if
short GRBs are not from compact star mergers, but from other
Type I progenitors (e.g. accretion induced collapses). It is hopeful 
that the upcoming Advanced LIGO \cite{smith09} experiment will be 
able to test these possibilities.  

Our first scenario (superposition between Type I and Type II) 
also demands that within bursts of massive star origin
(Type II), there could be two sub-types. One would need to 
find a reason to explain the apparent bimodal
distribution in the $T_{90}-$hardness space. It may be related to
the property of the progenitor star, or be related the process of
launching a relativistic jet. The duration of a GRB is related to
the duration of a relativistic jet that dissipates, which can be
shorter than the total time scale of accretion \citep{zhang09}.
Detailed models for short-duration Type II GRBs are called for
(e.g. Lazzati et al. 2009).

As shown, important and robust conclusions can be drawn about the 
nature of short/hard bursts from the current observations.  
Increasing the sample of short/hard bursts, especially those with 
redshift measurements and clear host galaxy associations, is of 
the greatest importance toward understanding the diverse underlying 
progenitors of these bursts and how we come to observe them.

\acknowledgments

This work partially supported by NSF under grant AST-0908362, and by
NASA under grants NNG05GB67G, NNX09AO94G, NNX08AE57A, NNX09AT66G, and
through the Nevada EPSCoR program (Nevada Space Grant).  E.T. was
supported by an appointment to the NASA Postdoctoral Program at the
GSFC, administered by Oak Ridge Associated Universities through a
contract with NASA.  We would also like to thank Chris Belczynski for
providing the redshift distribution data of NS-NS and NS-BH merger
events from his population synthesis code, and Rob Preece, Josh Bloom, 
Rachid Ouyed, and Amei Amei for helpful discussion and comments.

\begin{table}
\begin{center}
\caption{The $z$-known short/hard GRB sample}
\begin{tabular}{ccc}
\tableline
\tableline
GRB & z & $L_{\gamma,iso}^{peak}$ \\
name & redshift & $10^{50}$ erg s$^{-1}$ \\
\tableline
050509B & 0.2248 & $0.07^{+0.10}_{-0.05}$\\
050709 & 0.1606 & $5.4^{+0.67}_{-0.69}$\\
050724 & 0.2576 & $0.99^{+0.23}_{-0.10}$\\
060614 & 0.1254 & $1.39^{+0.13}_{-0.07}$\\
061006 & 0.4377 & $24.60^{+1.22}_{-0.77}$\\
050813 & 0.72 & $4.13 \pm 2.02$ \\
051221A & 0.5464 & $25.8 \pm 0.9$\\
060121 & 1.7/4.6 &$2445 \pm 162/33574 \pm 2226^a$ \\
060502B & 0.287 & $0.65 \pm 0.09$\\
060801 & 1.131 & $47.6^{+6.2}_{-1.6}$\\
061210 &  0.4095 & $21.5 \pm 1.4$\\
061217 & 0.8270 & $10.8 \pm 1.8$\\
070429B & 0.9023 & $24.6 \pm 3.8$\\
070714B & 0.9225 & $57.3 \pm 3.6$\\
070724A & 0.457 & $1.58^{+0.34}_{-0.14}$\\
071227 & 0.3940 & $3.34 \pm 0.49$\\
090426 & 2.6 & $171^{+24}_{-44}$ \\
090510 & 0.9 & $376^{+186}_{-172}$ \\
\tableline
\label{table}
\tablecomments{ Luminosities derived by author unless otherwise
  specified.  References for redshift measurements:
  \textbf{GRB050509B}:\cite{gehrels05}, \cite{bloom06}, \cite{ct05}; \textbf{GRB050709}:
  \cite{fox05},\cite{covino06},\cite{prochaska06}; \textbf{GRB050724}: \cite{berger05b}, \cite{prochaska06};
  \textbf{GRB060614}: \cite{dellavalle06}; \textbf{GRB061006}:
  \cite{berger07}; 
\textbf{GRB050813}: \cite{prochaska06}; 
\textbf{GRB051221A}: \cite{soderberg06}; \textbf{GRB060121}:
\cite z=1.7: {levan06}, \cite{berger07}, z=4.6: \cite{deugartepostigo06}  
\textbf{GRB060502B}: \cite{bloom07};
\textbf{GRB060801}:\cite{cucchiara06}, \cite{berger07}; 
\textbf{GRB061210}: \cite{berger07}; \textbf{GRB061217}:
\cite{berger07}; \textbf{GRB070429B}: \cite{cenko08};
\textbf{GRB070714B}: \cite{graham09}, \cite{cenko08}; \textbf{GRB070724A}:
\cite{cucchiara07}, \cite{berger09}, \cite{kocevski10}; \textbf{GRB071227}: \cite{davanzo09},\cite{berger09};
\textbf{GRB090426}: \cite{levesque09b},
\textbf{GRB090510}:\cite{rau09}, \cite{mcbreen10}; $^a$ We chose z=1.7 for this
analysis; $^b$ Derived from $\frac{E_{\gamma,iso}}{T_{90}}$. $T_{90}$:
\cite{palmer06}, $E_{\gamma,iso}$: \cite{hullinger06};}

\end{tabular}
\end{center}
\end{table}

\begin{sidewaystable}[h]
\scriptsize
\centering
\caption{Summary of merger models and statistical tests}
\begin{tabular}{ccccccc}
\hline
\hline
Model & LF parameters & $KS_z$ & $KS_L$ & $KS_t$ & BATSE LNLP & Swift LNLP$^a$ \\
  & $(\alpha_1,L_B, \alpha_2)$ & D-stat, Prob & D-stat, Prob & Prob &  T stat, P-value & T stat, P-value\\
\hline
1 Gyr $(\sigma=1.0)$ & (0.7,60,2.5) & 0.18, 0.69017 & 0.14, 0.91849 & 0.6339 & 2.12591, 0.04245 & -0.50769, 0.47997  \\
2 Gyr $(\sigma=1.0)$ & (0.42,40,2.5) & 0.14222, 0.90913 & 0.18, 0.69017 & 0.6275 &1.57805, 0.07254 & -0.55567, 0.49462  \\
3 Gyr $(\sigma=1.0)$ & (0.48,80,2.5) & 0.11333, 0.98782 & 0.12667, 0.96301 & 0.9513 & 2.60683, 0.02702 & 0.41429, 0.22775 \\
4 Gyr $(\sigma=1.0)$ & (0.19,40,2.5) & 0.15333, 0.85484 & 0.15778, 0.8301 & 0.7096 & 7.75112, 0.00041 & 1.25399, 0.10017  \\
5 Gyr $(\sigma=1.0)$ & (0.23,80,2.5) & 0.17556, 0.71954 & 0.18, 0.69017 & 0.4966 & 22.48737, 0 & 6.75693, 0.00090   \\
1 Gyr $(\sigma=0.3)$ & (0.93,80,2.5) & 0.19556, 0.58666 & 0.17778, 0.7049 & 0.4135 & 4.70011, 0.00469 & -0.36637, 0.4371   \\
2 Gyr $(\sigma=0.3)$ & (0.68,90,2.5) & 0.15333, 0.85484 & 0.16222, 0.80396 & 0.6873 & 1.44098 0.08312 & -0.69955, 0.53852  \\
3 Gyr $(\sigma=0.3)$ & (0.42,30,2.5) & 0.11556, 0.98491 & 0.15333, 0.85484 & 0.8419 & 2.67700, 0.02534 & -0.61568, 0.51296  \\
4 Gyr $(\sigma=0.3)$ & (0.35,50,2.5) & 0.12889, 0.957 & 0.14444, 0.89924 & 0.8606 &1.97297, 0.04921 & 0.25168, 0.26728  \\
5 Gyr $(\sigma=0.3)$ & (0.35,50,2.5) & 0.28, 0.17119 & 0.23333, 0.3608 & 0.0618 & 7.91458, 0.00036 & 5.38114, 0.00272 \\
Population synthesis & (0.19,80,2.5) & 0.14, 0.91849 & 0.12667, 0.96301 & 0.8845 & 45.97288, 0 & 3.72465, 0.01033  \\
Logarithmic & (0.08,80,2.5) & 0.15333, 0.85484 & 0.16444, 0.79044 & 0.6757 & 55.10492, 0 & 6.01050, 0.00164  \\
No delay & (1.15,80,2.5) & 0.19556, 0.58666 & 0.24667, 0.29602 & 0.17367 & 19.71989, 0  & 2.00273, 0.04781  \\
Twin & (0.14,30,2.5) & 0.20889	, 0.50096 & 0.19111, 0.61609 & 0.30864 & 2.45747, 0.03102 & -0.37388, 0.43936  \\
Mix 20 (PS)$^b$ & (0.24,80,2.5) & 0.16667, 0.77666 & 0.14, 0.91849 & 0.71336 & 32.62143, 0  & 3.77945, 0.00978 \\
Mix 50 (PS) & (0.2,90,2.5) & 0.15556, 0.84266 & 0.15111	, 0.86662 & 0.7302 &31.48321, 0 & 4.03421, 0.00798  \\
Mix 75 (PS) & (0.07,30,2.5) & 0.14444, 0.89924 & 0.12444, 0.96845 & 0.8709 &29.94587, 0  & 3.41307, 0.01332 \\
& (0.62,80,2.5) & 0.19333	, 0.60134 & 0.17556, 0.71954 & 0.4327 & 20.85024, 0 & 1.46488, 0.08117  \\
Mix 85 (PS) & (0.2,30,2.5) & 0.19556, 0.58666 & 0.11333, 0.98782 & 0.5795 & 17.84703, 0 & 1.30219, 0.09546  \\
Mix 90 (PS) & (0.1,30,2.5) & 0.273333, 0.192129 & 0.215556, 0.460233 & 0.0884 & 13.68442, 0 & 1.14002, 0.11223  \\
Mix 10 (Twin) & (0.61,90,2.5) & 0.11333, 0.98782 & 0.14, 0.91849 & 0.9073 &1.70715, 0.06384 & 3.43719, 0.01305  \\
Mix 20 (Twin) & (0.56,60,2.5) & 0.10889, 0.99239 & 0.10889, 0.99239 & 0.9848 & 2.56935, 0.02796 & -0.41484, 0.45173  \\
Mix 30 (Twin) & (0.33,20,2.5) & 0.16667, 0.77666 & 0.13556, 0.93559 & 0.7266 & 1.85675, 0.05511 & -0.32577, 0.42493   \\
Mix 40 (Twin) & (0.5,40,2.5) & 0.24667, 0.29602 & 0.19111, 0.61609 & 0.1824 & 1.60761, 0.07044 &  -0.14547, 0.37216  \\
\hline
\tablecomments{A summary of relevant merger delay models and the associated statistical tests with their test statistics and p-values.  Models that have not passed the $L$ and $z$ constraints are not included.  Our criteria for passing is at the 95\% level.  $^a$Comparison with the Swift short GRB sample with a truncation of $1.5 \rm~ph~cm^{-2}~s^{-1}$. $^b$ Mixing with the population synthesis model (PS).}
\end{tabular}
\end{sidewaystable}

\begin{figure}
\includegraphics[angle=0,scale=0.7]{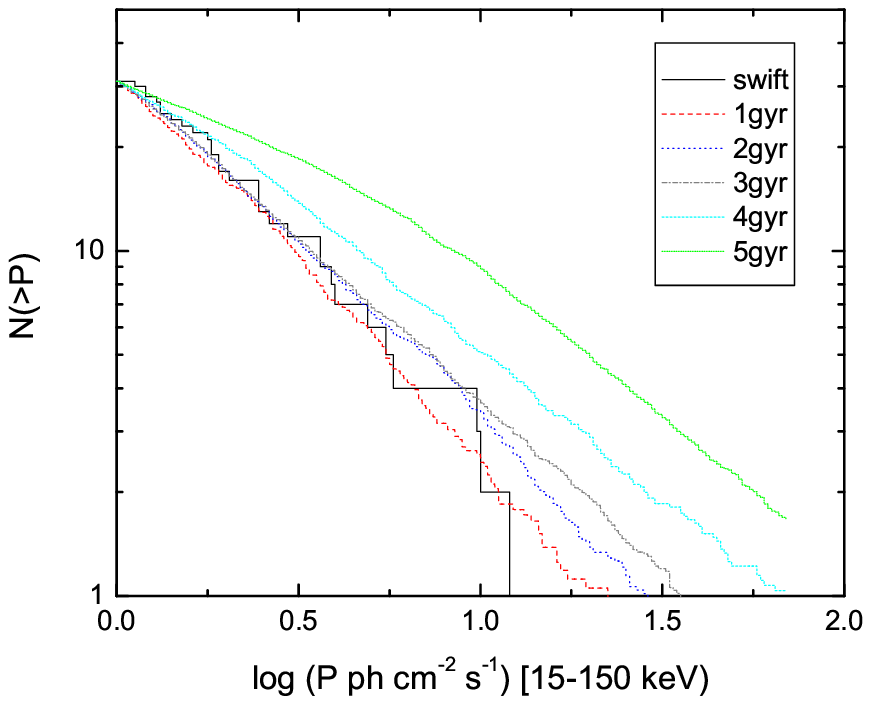}
\includegraphics[angle=0,scale=0.7]{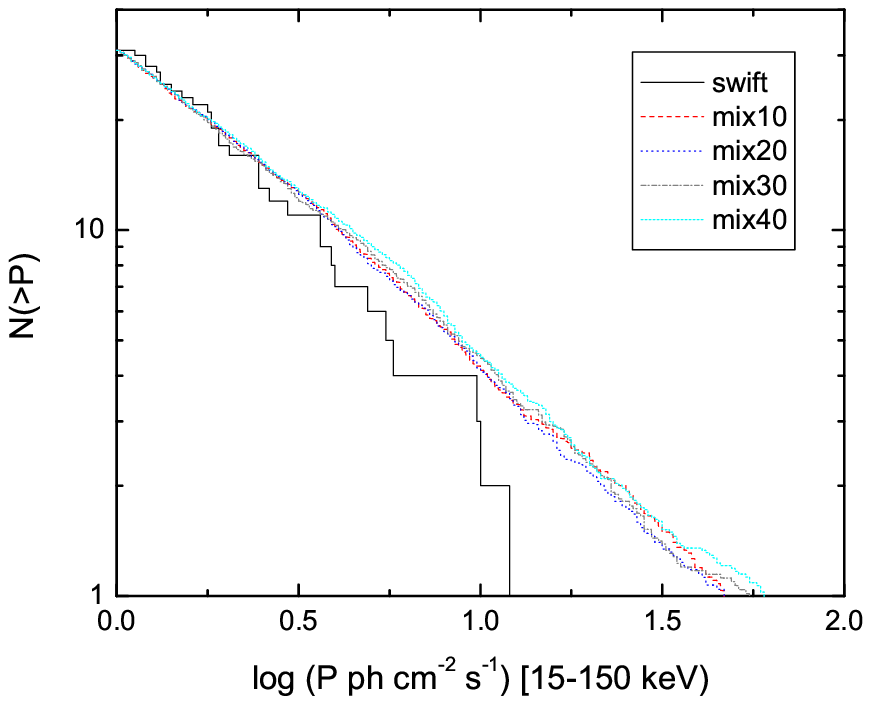}
\caption{$\log N-\log P$ distributions for the observed Swift sample and the simulated bursts in the Swift (15-150 keV) band.  Unlike the BATSE constraints, this test gives consistency for many more models and we present the most relevant ones here.  The first panel shows constant merger models ($\sigma=0.3$) and the second showing various mixed models with the ``twin'' population synthesis time delay model. }
\label{swiftlnlp}
\end{figure}

\begin{figure}
\includegraphics[angle=0,scale=0.7]{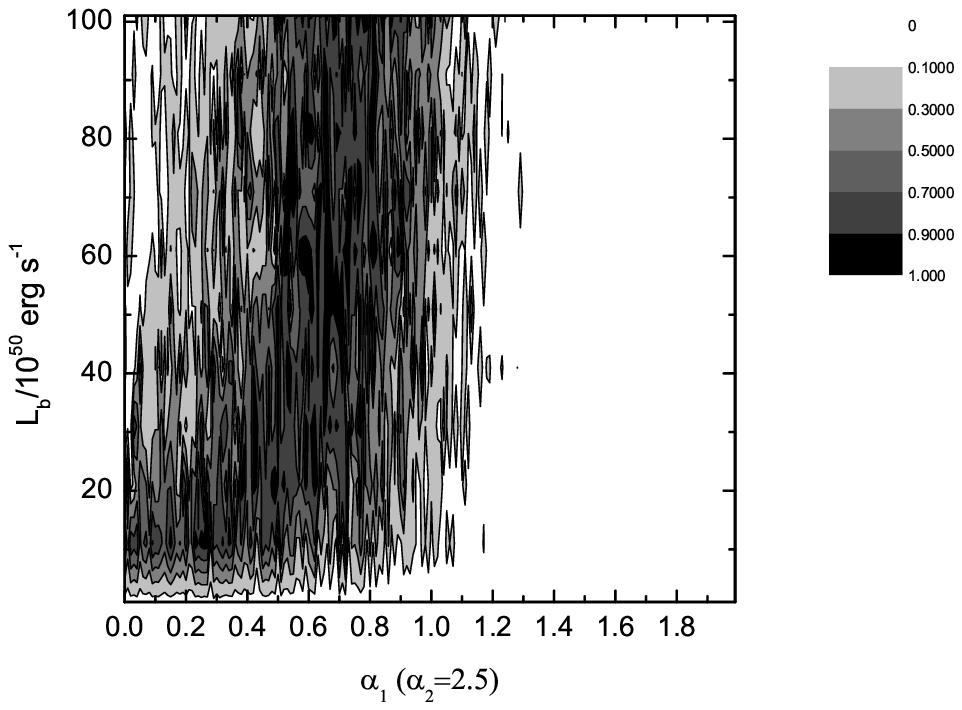}
\includegraphics[angle=0,scale=0.7]{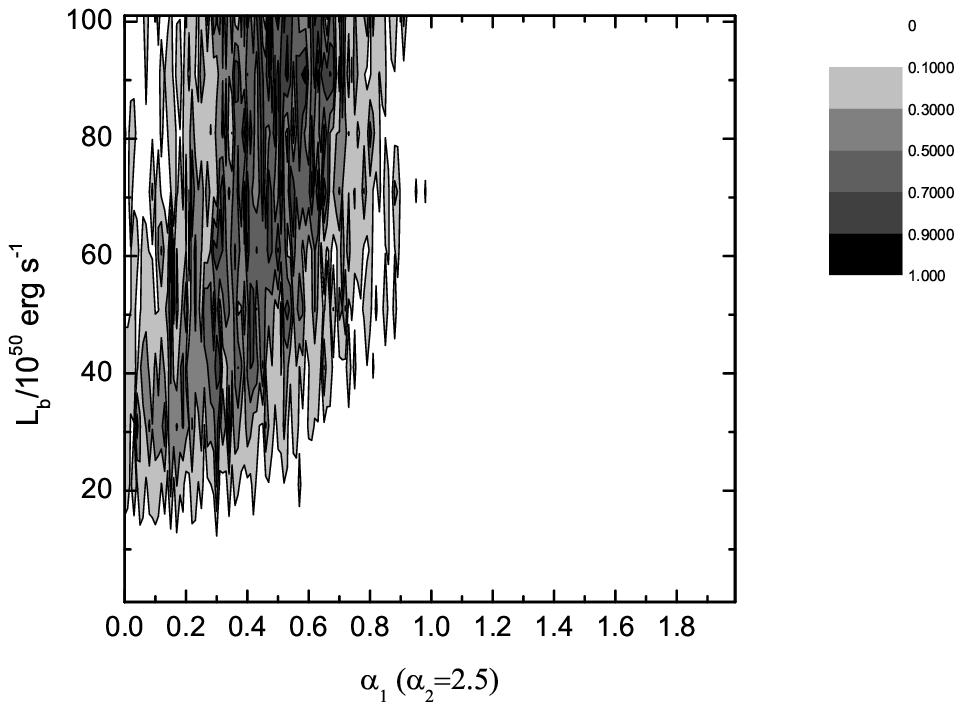}
\includegraphics[angle=0,scale=0.7]{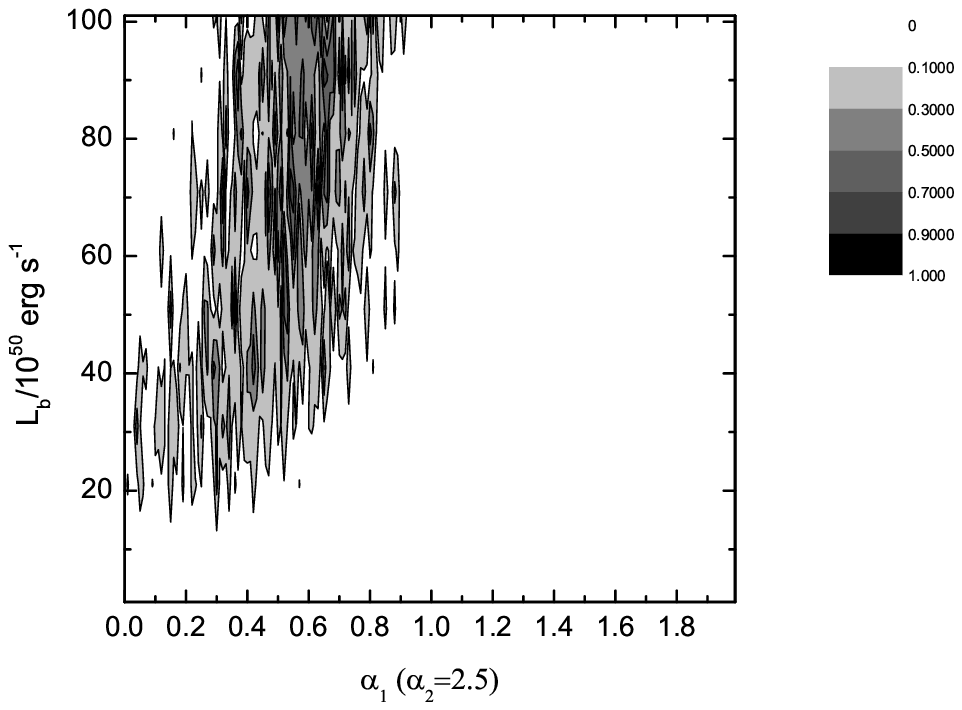}
\includegraphics[angle=0,scale=0.7]{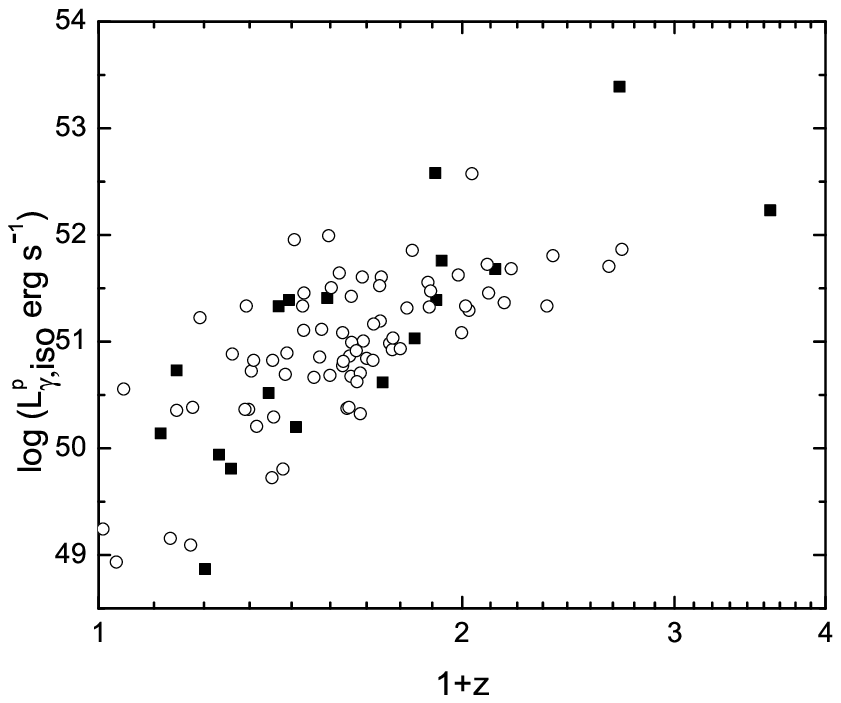}
\includegraphics[angle=0,scale=0.7]{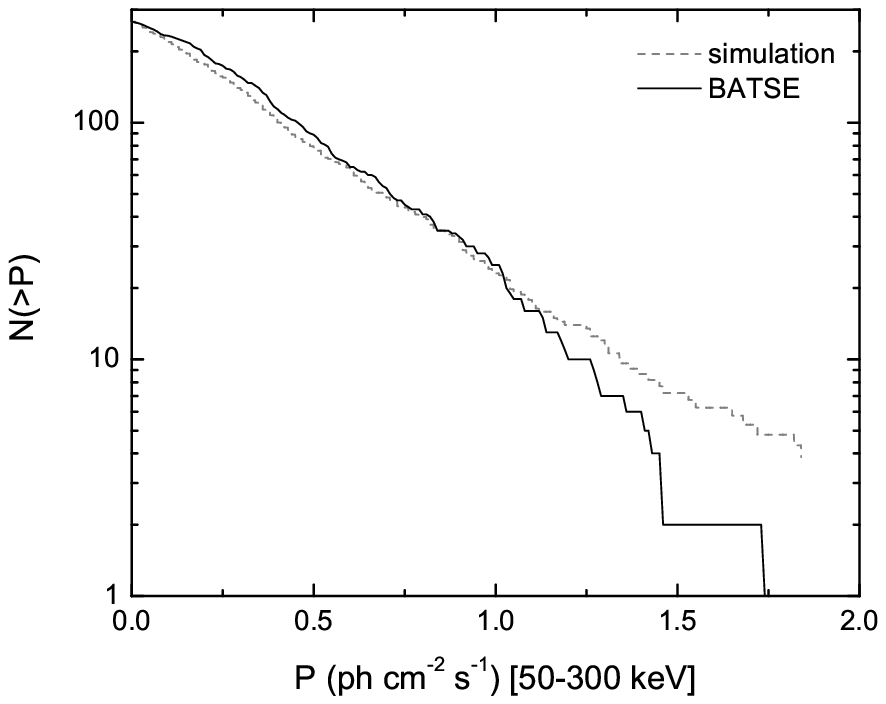}
\includegraphics[angle=0,scale=0.7]{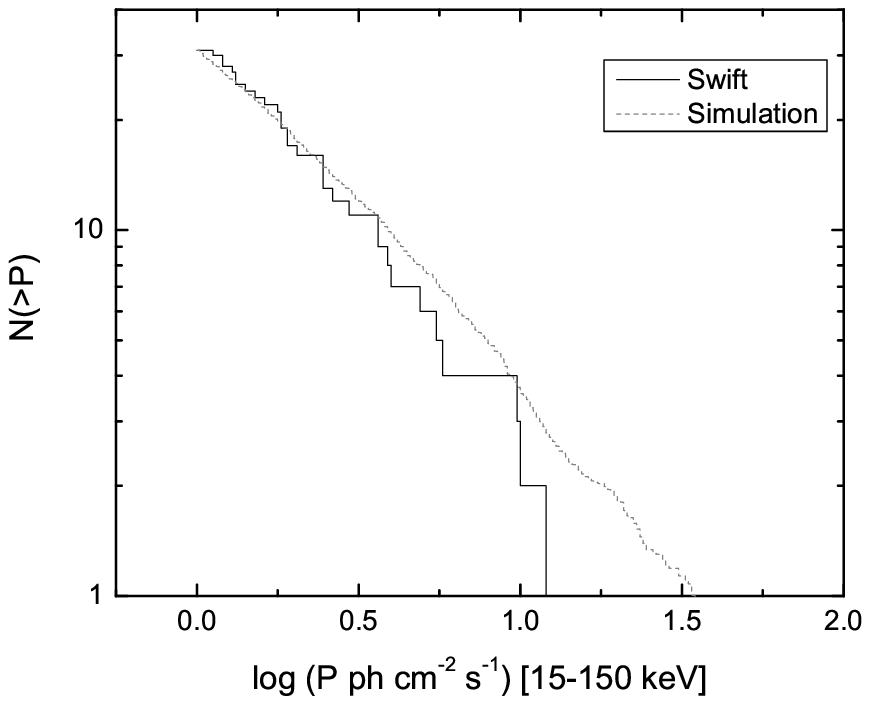}
\caption{Simulation results showing a distribution of short bursts
  that has a merger delay timescale of 2 Gyr ($\sigma=1.0$ Gyr), with luminosity
  function constrained by the $L-z$ data. The first three
  panels (a-c) are the $P_{KS,z}$, $P_{KS,L}$, $P_{KS,t}$ contours 
  (darker indicates higher KS probability). Panel (d) presents the
  simulated GRBs (open circles) with the best fit luminosity function 
  as compared with the data (solid dots) in the $L-z$ plane. Panel
  (e) and (f) shows the simulated $\log N-\log P$ (dashed line) as compared 
  with the BATSE (solid line) and Swift data, respectively.  Darker indicates higher KS probability and
  consistency with the observed $L$ and $z$ samples.}
\label{2gyr}
\end{figure}

\begin{figure}
\includegraphics[angle=0,scale=0.7]{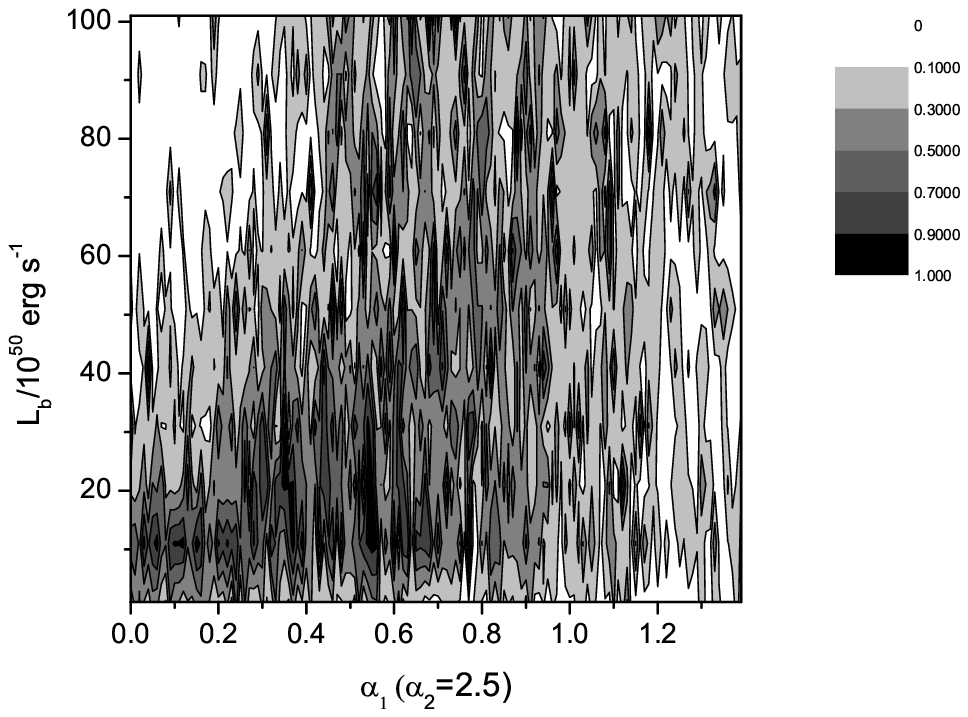}
\includegraphics[angle=0,scale=0.7]{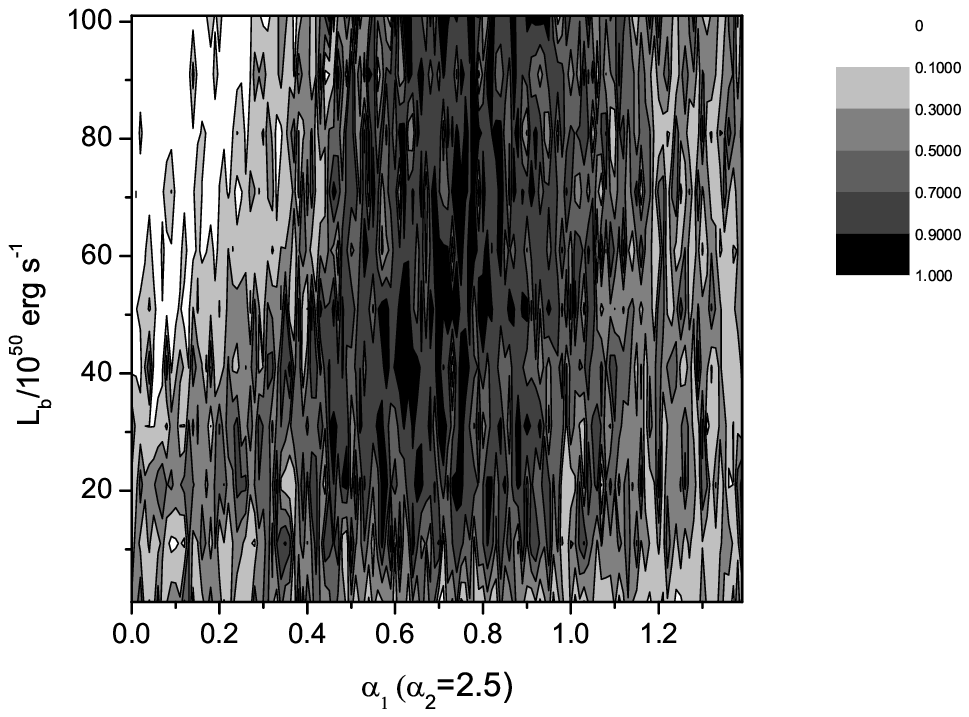}
\includegraphics[angle=0,scale=0.7]{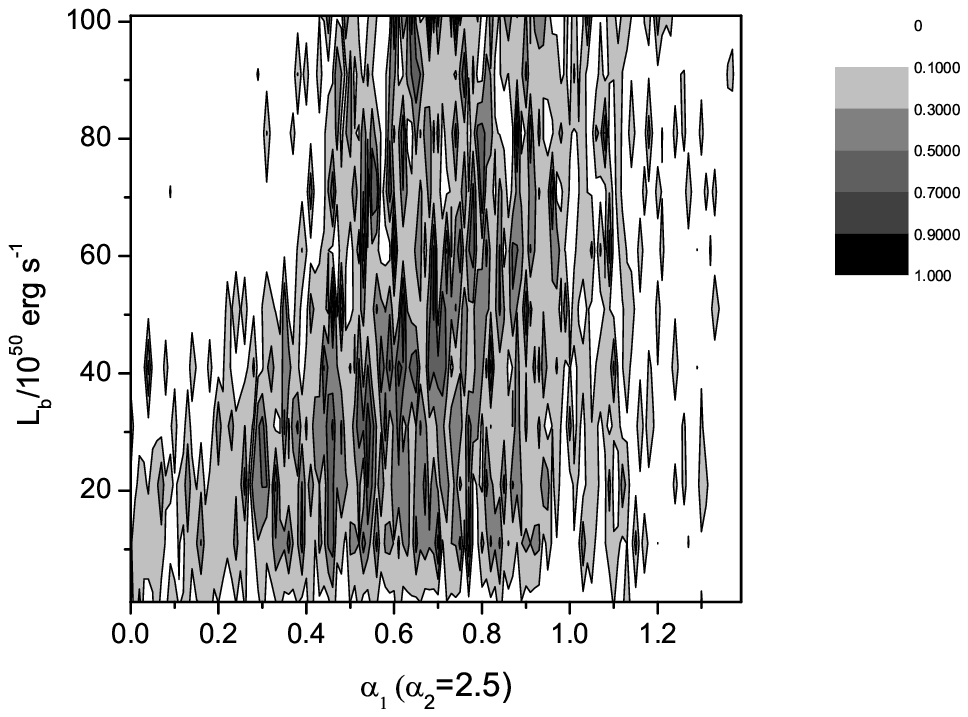}
\includegraphics[angle=0,scale=0.7]{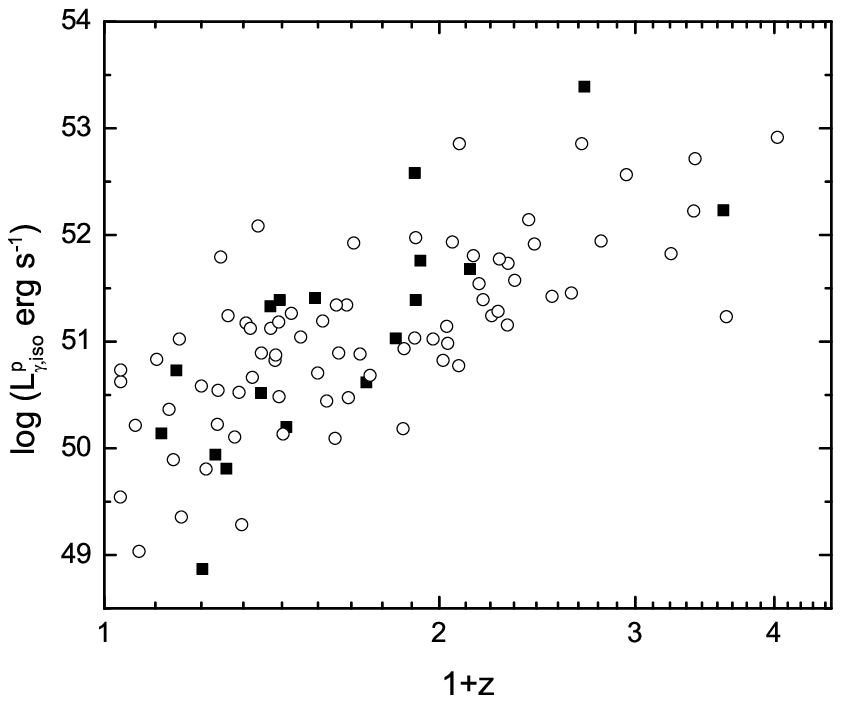}
\includegraphics[angle=0,scale=0.7]{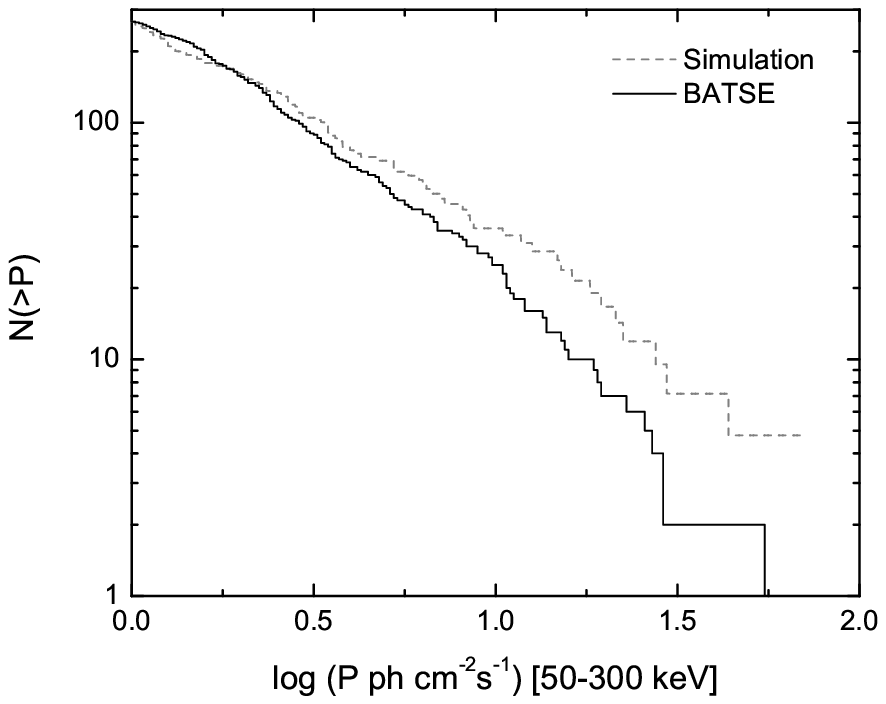}
\includegraphics[angle=0,scale=0.7]{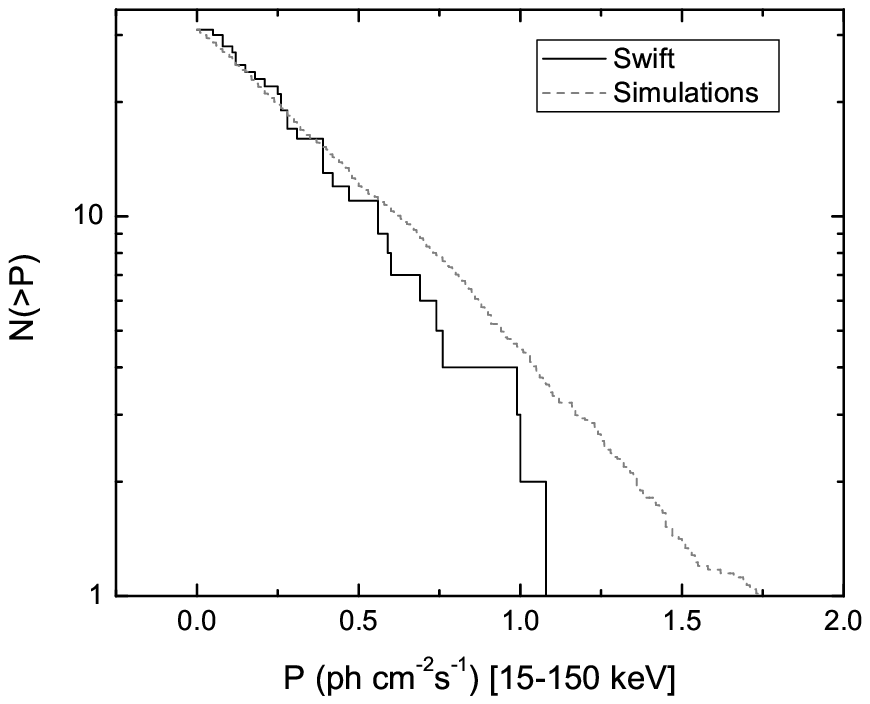}
\caption{A series of contours displaying the total KS probability,
  $P_{KS,t}$, for a model with a mix of 30\% of bursts following the star formation history and the Type II luminosity function and 70\% from the `twin' population synthesis model.  The first three
  panels (a-c) are the $P_{KS,z}$, $P_{KS,L}$, $P_{KS,t}$ contours 
  (darker indicates higher KS probability). Panel (d) presents the
  simulated GRBs (open circles) with the best fit luminosity function 
  as compared with the data (solid dots) in the $L-z$ plane. Panel
  (e) and (f) shows the simulated $\log N-\log P$ (dashed line) as compared 
  with the BATSE (solid line) and Swift data, respectively.  Darker indicates higher KS probability and
  consistency with the observed $L$ and $z$ samples.}
\label{mix30twin}
\end{figure}

\end{document}